\documentclass[
  aps,
  prl,
  twocolumn,
  superscriptaddress,
  nofootinbib]{revtex4-2}

\usepackage[normalem]{ulem}
\usepackage{xcolor}
\usepackage[
    colorlinks=true,
allcolors=blue!70!black
]{hyperref}
\usepackage{amsmath,amssymb}
\usepackage{mathrsfs}
\usepackage{lineno}

\usepackage{graphicx,amsmath,amsfonts}
\usepackage{amssymb}
\usepackage{amsthm}
\usepackage[T1]{fontenc}

\newcommand{\vers}[1]{\hat{\mathbf{#1}}}

\usepackage{xcolor}
\usepackage{amsfonts}
\usepackage{amsmath}
\usepackage{amssymb}
\usepackage{mathtools}

\def\XXint#1#2#3{{\setbox0=\hbox{$#1{#2#3}{\int}$}
     \vcenter{\hbox{$#2#3$}}\kern-.5\wd0}}

\renewcommand{\vec}[1]{\mathbf{#1}}

\newcommand{\V}{\Omega}
\newcommand{\dV}{\, \mathrm{d}\V}
\newcommand{\dS}{\, \mathrm{d}S}
\newcommand{\sca}{\mathrm{s}}
\newcommand{\abs}{\mathrm{a}}
\newcommand{\ext}{\mathrm{e}}
\newcommand{\inc}{\mathrm{i}}

\newcommand{\ed}{\Omega}

\newtheorem{corollary}{Corollary}

\newcommand{\zsp}{\zeta_{\Omega}^{\mathrm{sp}}}

\begin{document}

\title{Bode--Fano Limits to Broadband Absorption by Small Particles}

\author{Emanuele Corsaro}
\affiliation{ Department of Electrical Engineering and Information Technology, Universit\`{a} degli Studi di Napoli Federico II, via Claudio 21, Napoli, 80125, Italy}
\affiliation{Photonics Initiative, Advanced Science Research Center, City University of New York, New York,
New York 10031, USA}
\author{Andrea Alù }
\email[]{aalu@gc.cuny.edu}
\affiliation{Photonics Initiative, Advanced Science Research Center, City University of New York, New York,
New York 10031, USA}
\affiliation{Physics Program, Graduate Center of the City University of New York, New York, New York 10016, USA}
\author{Carlo Forestiere}
\email[]{carlo.forestiere@unina.it}
\affiliation{ Department of Electrical Engineering and Information Technology, Universit\`{a} degli Studi di Napoli Federico II, via Claudio 21, Napoli, 80125, Italy}

\begin{abstract}

Subwavelength nanostructures can absorb light efficiently at resonance, but causality and passivity fundamentally limit the bandwidth over which strong absorption can be maintained. 
 We derive Bode–Fano-type limits for electrically small scatterers dominated by a single mode, assuming linear, time-invariant materials with rational dispersion. The derivation maps electromagnetic absorption onto an equivalent impedance-matching problem and provides a constructive method for synthesizing material dispersion that approaches these bounds while satisfying the same physical constraints.
Our results guide the design of passive nanoparticles for broadband absorption, with potential applications in nanoscale light harvesting and dissipation.
\end{abstract}

\maketitle

{ In linear, time-invariant electromagnetic systems, causality constrains the analytic structure of response functions and underlies Kramers--Kronig dispersion relations. These analytic properties, together with the positivity constraints imposed by passivity, provide a basis for fundamental performance bounds.} Beyond their theoretical value, such limits offer reliable signposts for technology: they quantify unavoidable trade-offs, and they provide an objective reference to assess whether a given structure is near-optimal or whether further improvement is possible by changing materials, constraints or the operating regime \cite{chao_physical_2022}.

A relevant example is impedance matching in electrical network theory, which aims to suppress reflections and maximize power transfer towards a load. In this setting, Bode--Fano integral bounds limit the bandwidth over which a passive, lossless, linear, time-invariant matching network can deliver a prescribed minimum fraction of the available power from a resistive source to a given load with both resistive and reactive components \cite{bode_network_1945,fano_theoretical_1947}. Analogous restrictions arise for electrically small antennas, where unavoidable stored energy imposes lower bounds on the radiation quality factor \cite{chu_physical_1948,wheeler_fundamental_1947,harrington_effect_1960,gustafsson_physical_2012}.

In the context of light--matter interactions, related ideas have motivated a broad search for fundamental limits on scattering, extinction, and absorption properties of small particles. Single-frequency bounds derived from energy conservation quantify how the material susceptibility limits far- and near-field observables at a given frequency~\cite{miller_fundamental_2014,miller_fundamental_2016,kuang_maximal_2020}. In parallel, sum rules connect the global response function integrated over all frequencies to the nanoparticle geometry and the static or high-frequency material response{, yielding} bandwidth--magnitude trade-offs that are independent of detailed device implementations \cite{sohl_physical_2007,sohl_physical_ostacles_2007,molesky_global_2020,gustafsson_upper_2020}, and power--bandwidth limits for the local density of states and radiative heat transfer \cite{shim_fundamental_2019}.

{ These sum rules constrain weighted integrals of the response over all frequencies. By themselves, they do not explicitly capture how the bandwidth bound tightens as the prescribed absorption level approaches its modal maximum.}

{ Addressing this trade-off requires additional constraints on the
admissible spectral response, motivating a Bode--Fano formulation
in terms of the mismatch from maximum absorption.
Related bandwidth limitations have been derived in other
electromagnetic settings, including Rozanov-type bounds for planar
absorbers~\cite{rozanov_ultimate_2000} and Bode--Fano-inspired bounds
for passive cloaking~\cite{monticone_invisibility_2016}.}

Here, we derive Bode--Fano-type integral constraints for electrically small scatterers whose response is dominated {by a single mode, assuming passive, causal, linear, time-invariant materials with rational dispersion.} These constraints directly limit the bandwidth over which a prescribed absorption level can be maintained. { We illustrate the theory for an electric-dipole mode. The Supplemental Material~\cite{SuppMat} provides the corresponding derivation for an isolated magnetic-dipole mode.}

{ Using the radiation and material impedances introduced in Ref.~\cite{forestiere_first-principles_2024}, we map single-mode electromagnetic absorption onto an impedance-matching problem that separates geometry from material dispersion. We establish that both impedances are positive-real functions, enabling the application of classical passive matching theory.}
As a result, we obtain Bode--Fano constraints directly linked to
the geometry of the scatterer. Beyond { establishing these} limits the same correspondence
provides a constructive framework for { synthesizing} the  material dispersion of { a particle of prescribed geometry so as to approach these integral bounds, while preserving passivity and causality}.


\begin{figure}
    \centering
    \includegraphics[width=0.95\linewidth]{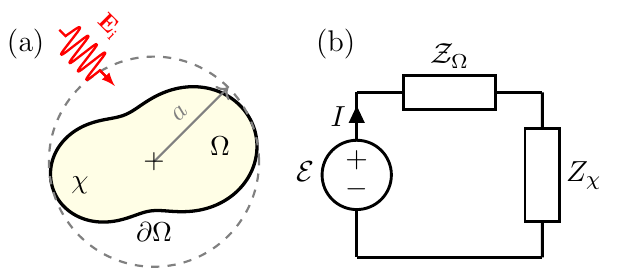}
    \caption{(a) A temporally dispersive object { occupying the domain $\Omega$ enclosed within the ball $B_a$ of radius $a$}. The object is illuminated by a time-harmonic electric field $\vec{E}_\inc$. (b) Equivalent circuit model, mapping the response of the polarization current density induced by the external field $\mathbf{E}_\inc$ within the dipolar approximation. }
    \label{fig:intro}
\end{figure}

\paragraph{Formulation of the Problem---} Consider an object made of a linear, homogeneous, and time-invariant material with electric susceptibility $\chi$, occupying a bounded domain $\Omega$  embedded in vacuum. The object is illuminated by a time-harmonic incident electric field $\mathbf{E}_{\inc}$ at angular frequency $\omega$, using the $e^{j\omega t}$ convention.  Let $B_a $ denote the smallest circumscribing { ball} of radius $a$ such that $\Omega\subseteq B_a$ [Fig. \ref{fig:intro}(a)].  

In the volume-integral formulation~\cite{hanson_operator_2002,mishchenko_electromagnetic_2014}, { the}  induced polarization current density $\mathbf{J}$, supported in $\Omega$, { satisfies $(Z_\chi \mathbb{I} + \mathcal L_\Omega)\vec{J}= a^{-1} \mathbf{E}_{\inc}$. Here, $Z_\chi :=(j\omega \varepsilon_0a \chi)^{-1}$ is the \textit{material impedance} \cite{forestiere_first-principles_2024}, $\mathbb{I}$ is the identity operator, and $\mathcal{L}_\Omega$ is the free-space radiation operator, defined in the Supplemental Material~\cite{SuppMat},  mapping a polarization current density onto the electric field it radiates in vacuum, up to the constant factor \(-a\).}
Following Ref. \cite{forestiere_first-principles_2024}, we expand $\vec{J}$ in terms of the polarization current density modes $\{\vec{j}_h\}_{h=1}^\infty$ as $\vec{J}=a^{-2}\sum_h I_h \vec{j}_h$.
{ These modes are defined through the eigenvalue problem $\mathcal{L}_\Omega \vec{j}_h = \zeta_{\Omega,h} \vec{j}_h$ with  $\vec{r}\in\Omega$ \cite{forestiere_first-principles_2024,SuppMat}. The associated eigenvalue $\zeta_{\Omega,h}$ is the \textit{radiation impedance}~\cite{forestiere_first-principles_2024} of the $h$th mode. Importantly, both $\mathbf{j}_h$ and $\zeta_{\Omega,h}$ are determined by the scatterer geometry {  and the operational frequency} but are independent of the material susceptibility. This material independence is central to the formulation of this Letter, as it enables a clean separation between geometrical and material contributions, represented by the radiation impedance $\zeta_{\Omega,h}$ and the material impedance $Z_\chi$, respectively.}

We assume that { (H1)} the response is dominated  by an isolated electric dipole mode  ($h=1$); { (H2) the material is passive and causal, with a rational frequency-dependent susceptibility; and (H3) the particle is electrically small, i.e., $\omega a/c_0 \ll 1$, in the band of interest, and thus we adopt
the rational positive-real small-particle approximation to $\zeta_\Omega(s)$ specified in the Supplemental Material. Under (H1),} the scattering problem { reduces} to the equivalent single-mode circuit in Fig.~\ref{fig:intro}(b) \cite{forestiere_first-principles_2024}.  { The equivalent source $\mathcal{E}$ represents the excitation, while the material impedance $Z_\chi=R_\chi+jX_\chi$ and the radiation impedance of the electric dipole mode $\zeta_{\ed}=\mathcal{R}_{\ed}+j\mathcal{X}_{\ed}$ are connected in series~\cite{forestiere_first-principles_2024}.} The driving-point admittance $Y = G+jB$ is given by $Y=(Z_\chi + \zeta_\ed)^{-1}$.  { Accordingly}, the extinguished, scattered, and absorbed powers are  \cite{corsaro_mie_2026,SuppMat}
\begin{equation}
P_{\ext} = \frac{|\mathcal{E}|^2}{2} G; \,\,\,
P_{\sca} = \frac{|\mathcal{E}|^2}{2} \mathcal{R}_{\ed} |Y|^2; \,\,\,
P_{\abs} = \frac{|\mathcal{E}|^2}{2} R_\chi |Y|^2.
\label{eq:powers}
\end{equation}
{ The radiation impedance is analytic in the open right half plane $\mathbb{C}_+$ \cite{forestiere_first-principles_2024}. Moreover,} the Laplace-domain  Poynting theorem establishes that $\Re\{\zeta_{\ed}(s)\}\ge0$ in $\mathbb{C}_+$ and that $\zeta_{\ed}(s)$ is real for positive real $s$~\cite{SuppMat}. Together, these properties make $\zeta_{\ed}$ a \textit{positive-real function} (PRF)~\cite{brune_synthesis_1931}. 

Since the material response is passive and causal, the material impedance \(Z_\chi\) is also a PRF, rational by hypothesis (H2). Because the sum of two PRFs is also a PRF, and the reciprocal of a PRF is again a PRF, it follows that $Y$ is positive-real. 
{ It therefore admits the  integral representation in $\mathbb{C}_+$ \cite{cauer_poisson_1932} 
\begin{equation}
    Y(s) = sC_{\infty} + \frac{2s}{\pi}\int_0^\infty \frac{G(\omega)}{s^2+\omega^2} \mathrm{d}\omega,
    \label{eq:KK_relation}
\end{equation}
where $G(\omega)=\Re \{ Y(j\omega) \}$ and
$C_{\infty}=\lim_{s\to\infty}Y(s)/s$. 

At the origin, $\zeta_{\ed}$ has a simple pole with residue $C_{\ed,0}^{-1}>0$, while $Z_\chi$ has at most a simple pole with nonnegative residue $C_{\chi,0}^{-1}:=\lim_{s\to0^+}sZ_\chi(s)$. Since these residues add, $Y(s)=sC_0+O(s^2)$, where the static capacitance satisfies
$ C_0:=\lim_{s\to0^+}{Y(s)}/{s} =(C_{\ed,0}^{-1}+C_{\chi,0}^{-1})^{-1} \le C_{\ed,0}.
$


Dividing Eq.~\eqref{eq:KK_relation} by $s$ and taking $s\to0^+$ gives~\cite{bode_network_1945,gustafsson_sum_2010,SuppMat}}
\begin{equation}
\int_{0}^{\infty} G(\omega)\,
\frac{\mathrm{d}\omega}{\omega^{2}}
=
\frac{\pi}{2}\left(C_0-C_\infty\right)
\leq
\frac{\pi}{2} C_{\ed,0},
\label{eq:Sum_Rule_ext}
\end{equation}
where the final inequality also uses $C_\infty\ge0$.

For a fixed geometry, the upper bound in Eq.~\eqref{eq:Sum_Rule_ext} is attained when $C_{\chi,0}^{-1}=0$ and $C_\infty=0$, so that $C_0=C_{\ed,0}$. The first condition excludes a pole of $Z_\chi$ at the origin and gives $C_0=C_{\ed,0}$. { A sufficient condition for the second is the high-frequency susceptibility  asymptotics~\cite{altarelli_superconvergence_1972} $\chi(s)=\chi_{\infty}s^{-2} +O\!\left(s^{-3}\right)$, for $s\to\infty $, with $\chi_{\infty}>0$. Indeed, this condition implies that the material impedance  is inductive at $s=\infty$,  $Z_{\chi}(s)= sL_{\chi,\infty} + O(1)$ with
$L_{\chi,\infty}=(\varepsilon_{0}a\chi_{\infty})^{-1}>0$, so that
$Y(s)=O(s^{-1})$ and hence $C_{\infty}=0$.}
 Both conditions hold for a series $RL$ material impedance, corresponding to the Drude susceptibility introduced below.

Equation \eqref{eq:Sum_Rule_ext} is closely related to the extinction sum rules of Refs.~\cite{purcell_absorption_1969,sohl_physical_ostacles_2007,bohren_absorption_1998}. Our derivation, however, has different basis: the sum rules in
Refs.~\cite{purcell_absorption_1969,sohl_physical_ostacles_2007,bohren_absorption_1998}
were formulated for the extinction cross section and rely on causality arguments whose validity has been questioned in Ref.~\cite{mishchenko_broadband_2008}. By contrast,
Eq.~\eqref{eq:Sum_Rule_ext} is obtained as a sum rule for the driving-point admittance of the circuit model. The analyticity of $Y$ follows from the positive-real character of the material and radiation impedances. 

By fixing the total available spectral area, this sum rule implies an overall bandwidth limitation on the extinction response (see e.g. Ref.~\cite{sohl_physical_2007}). { Since $P_\abs\le P_\ext=P_\abs+P_\sca$, it also provides an absorption bound. This bound can be loose, however, because it allocates the entire extinction budget to absorption without accounting for the scattering contribution.}

However, { the sum rule alone does not capture a crucial limitation: the bandwidth over which absorption remains above a prescribed level must shrink as that level approaches the modal maximum.  We derive this additional restriction through Bode--Fano-type constraints that exploit the analytic structure of the material and radiation  impedances \cite{bode_network_1945,fano_theoretical_1947, youla_new_1964}.

{
To this end, we introduce the reflection coefficient \cite{youla_new_1964}
\begin{equation}
    \Gamma(s) := \frac{Z_{\chi}(s) - \zeta_\Omega(-s)}{Z_{\chi}(s) + \zeta_\Omega(s)}.
    \label{eq:Gamma}
\end{equation}
}

In terms of $\Gamma(j\omega)$, the extinguished, scattered and absorbed powers, normalized to
$P_{\mathrm{av}}=|\mathcal{E}|^2/(8\mathcal{R}_\ed)$, read \cite{SuppMat}:
\begin{equation}
\frac{P_\ext}{P_{\rm av}} = 2(1-\Re\{\Gamma\}); \,\,\, 
\frac{P_\sca}{P_{\rm av}}=|1-\Gamma|^2;
\,\,\,
\frac{P_{\abs}}{P_{\rm av}}=1-|\Gamma|^2.
\label{eq:powers_gamma}
\end{equation}
Passivity implies $|\Gamma(j\omega)|\leq 1$. Hence, from
Eq.~\eqref{eq:powers_gamma}, it follows that
$P_{\abs}/P_{\rm av}\leq 1$, $P_\ext /P_{\rm av} \leq 4 $, $P_\sca /P_{\rm av} \leq 4 $; { yielding} upper bounds on the maximum level of the corresponding cross-sections, analogous to the single-frequency bounds of Ref.~\cite{miller_fundamental_2016}.

At this point, the electromagnetic scattering problem takes the form of an impedance-matching problem, a topic that has been studied for decades in network theory \cite{fano_theoretical_1947, pozar_microwave_2012}. From Eq. \eqref{eq:powers_gamma}, it follows that maximum absorption, i.e., \(P_{\abs}=P_{\rm av}\), is attained at the frequencies for which \(Z_\chi\) is \textit{conjugate matched}
to  \(\zeta_{\ed}\), namely when $\Gamma=0$, corresponding to maximum power transfer from the source to \(Z_\chi\).  At such frequencies, the real and imaginary parts of $Z_\chi$ and $\zeta_\ed$ satisfy $R_\chi =\mathcal R_{\ed}$ and $X_\chi +\mathcal X_{\ed} =0 $. These conditions identify the resonant frequencies at which the absorption response is maximum ($P_\abs = P_{\rm av}$). The Bode--Fano integral relations for $\Gamma$, derived in the Supplemental Material~\cite{SuppMat}, constrain how closely this absorption limit can be approached over a finite frequency band.
}

To derive these relations, we must account for the zeros and singularities of $\Gamma$ in $\mathbb C_+$. Both $Z_\chi(s)$ and $\zeta_\Omega(s)$
are analytic there, and their positive-real character ensures that their sum does not vanish. Thus, the singularities of $\Gamma$ arise solely from the reflected radiation
impedance $\zeta_\Omega(-s)$. Under hypothesis~(H3), we adopt the rational small-particle approximation to $\zeta_\Omega(s)$. Besides its simple pole at the origin, this approximation has exactly one pole, which is simple and located on the negative real axis. The corresponding reflection coefficient is therefore meromorphic in $\mathbb C_+$, with a single simple real pole there. Under these hypotheses, the Bode-Fano integral
relations then yield \cite{SuppMat}
\begin{equation}
\int_0^\infty
\ln\frac{1}{|\Gamma(j\omega)|}
\frac{\mathrm d\omega}{\omega^4}
\leq
\frac{2\pi}{3}\left(\frac{a}{c_0}\right)^3,
\qquad
\forall\,\Omega\subseteq B_a.
\label{eq:bode-fano}
\end{equation}
Among all geometries contained in a sphere of radius $a$, the sphere itself maximizes the geometry-dependent upper bound on the electric-dipole Bode--Fano integral.

{ The framework can be extended to other isolated modes by considering the appropriate modal radiation impedance. The Supplemental Material~\cite{SuppMat} provides the derivation for an isolated magnetic-dipole mode.

To make the bandwidth--matching trade-off { imposed by \eqref{eq:bode-fano}} explicit, we consider an absorption
response satisfying
\(P_{\abs}/P_{\rm av}\geq 1-\delta\) throughout the band \(\mathcal{B} =[\omega_1,\omega_2]\). We  define the bandwidth $\Delta\omega=\omega_2-\omega_1$ and geometric center frequency $\omega_0=\sqrt{\omega_1\omega_2}$. Since
\(P_{\abs}/P_{\rm av}=1-|\Gamma|^2\), this requires
\(|\Gamma|\leq\Gamma_{\max}:=\sqrt{\delta}\) throughout the band.  Equation~\eqref{eq:bode-fano} then gives
\begin{equation}
\ln\!\left(\frac{1}{\Gamma_{\max}}\right)
\frac{\Delta\omega}{\omega_0}
\leq
{\frac{\pi}{Q_{\min}}},
\label{eq:bode-fano-tradeoff}
\end{equation}
where $Q_{\min}=(3/2)(\omega_0 a/c_0)^{-3}$ is the minimum quality factor of a small electric-type resonator, attained by the electric-dipole mode of a sphere \cite{chu_physical_1948,wheeler_fundamental_1947,harrington_effect_1960,gustafsson_physical_2012, pascale_lower_2023}.
Thus, for fixed { radius of the circumscribing sphere}, increasing the required absorption level, or equivalently decreasing \(\Gamma_{\max}\),  tightens the bandwidth bound.

Equations~\eqref{eq:bode-fano} and~\eqref{eq:bode-fano-tradeoff} rely on the rational small-particle approximation to the radiation impedance. In the full-wave problems, modal coalescences may generate branch points in the analytic continuation of $\zeta_\Omega(s)$ into the open left half-plane. Through the reflected impedance $\zeta_\Omega(-s)$, these singularities can be inherited by $\Gamma(s)$ in Eq.~\eqref{eq:Gamma}, producing additional branch-cut contributions to the Bode--Fano integral relations \cite{SuppMat}. These additional terms depend on the material impedance through $\Gamma(s)$, { making the derivation of a
material-independent bound for the full-wave
problem nontrivial.}

\begin{figure}
    \centering
    \includegraphics[width=1\linewidth]{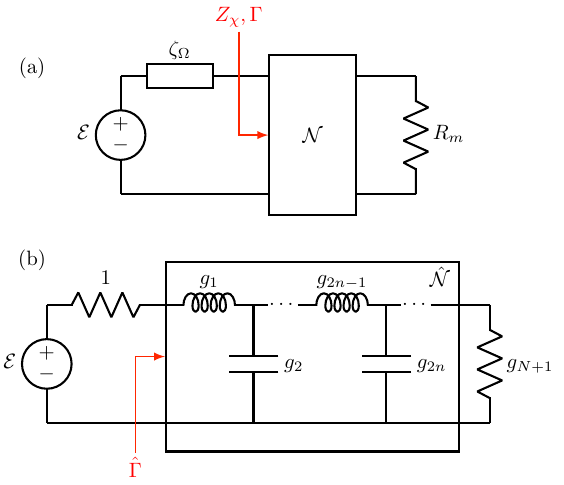}
    \caption{(a) Darlington representation of $Z_\chi$. (b) Low-pass prototype in the $\hat{s}$-domain.}
    \label{fig:lpproto}
\end{figure}


\paragraph{Dispersion Synthesis---}
To construct material dispersion consistent with passivity and causality, we invoke Darlington's theorem \cite{darlington_synthesis_1939}: any rational positive-real impedance, such as $Z_\chi$, can be realized as the input
impedance of a lossless two-port network $\mathcal N$
terminated by an arbitrary resistor $R_m>0$. Accordingly, the circuit in Fig.~\ref{fig:intro}(b)
admits the representation in Fig.~\ref{fig:lpproto}(a).

We first identify the lowest-order passive material
impedance $Z_\chi$ capable of resonating with the electric-dipole radiation impedance $\zeta_{\ed}$.

{ Since $\zeta_\Omega$ is ohmic-capacitive} 
\cite{forestiere_first-principles_2024,corsaro_mie_2026}, the simplest such realization is a series $RL$ circuit,  $Z_\chi(s)=R_m+sL_m$}.  { For a prescribed resonance frequency $\omega_0$}, the zero-reactance condition { yields $L_m = -\mathcal{X}_\Omega(\omega_0)/\omega_0$, and $\zeta_\Omega(j\omega_0) + Z_\chi(j\omega_0) = R_s+R_m$, where $R_s= \mathcal{R}_\Omega(\omega_0)$, and $R_m = R_\chi(\omega_0)$} \cite{SuppMat}.
In this minimal realization, conjugate matching,
and hence maximum modal absorption at $\omega_0$,
is achieved when $R_m=R_s$.
{ Near resonance,  we approximate the radiation resistance
by $R_s$ and match the total reactance to first order
in $\omega-\omega_0$.} The normalized driving-point
impedance takes the form
\begin{equation}
    \frac{Z_\chi(s)+\zeta_{\ed}(s)}{R_s}
    \simeq
    1+r_m+
    g_1\frac{s^2+\omega_0^2}{\beta s},
    \label{eq:driving_point_impedance}
\end{equation}
where
$r_m=R_m/R_s$,
$g_1=
\beta[\mathcal X_\ed'(\omega_0)+L_m]/(2R_s)$,
and $\beta>0$ is a frequency-scaling parameter. This motivates the dimensionless low-pass
frequency transformation
\cite{pozar_microwave_2012,su_analog_1996}
{
\begin{equation}
    \hat{s}
    =
    \frac{s^2+\omega_0^2}{\beta s},
    \label{eq:Lp2Bp}
\end{equation}
}
where $\hat{s}=\hat{\sigma}+j\hat{\omega}$
denotes the transformed low-pass variable. { Indeed, in the $\hat{s}$-domain, the right-hand side of Eq. \eqref{eq:driving_point_impedance} becomes
$1+r_m+g_1\hat{s}$, corresponding to a normalized low-pass network consisting of a unit source resistance, a series inductance $g_1$, and a terminating resistance $r_m$. The synthesis can therefore be formulated as a standard
low-pass filter-design problem \cite{pozar_microwave_2012,su_analog_1996,van_valkenburg_network_1974}}. Moreover, this transformation maps the resonant frequency $\omega_0$ into $\hat{\omega}=0$, while the bandwidth $\Delta\omega$ is mapped to the low-pass cutoff frequency $\hat{\omega}_c=\Delta\omega/\beta$
\cite{SuppMat}.

Having established the minimal realization, we next
consider higher-order realizations of the material
impedance $Z_\chi$ that allow the material
dispersion to be tailored to broaden the absorption bandwidth at fixed geometry while preserving causality and
passivity.
For a meaningful comparison with the minimal case,
we retain the same resonance frequency $\omega_0$
and material resistance at resonance:
$\mathcal X_{\ed}(\omega_0)+X_\chi(\omega_0)=0$
and $R_\chi(\omega_0)=R_m$.
These conditions preserve the absorption level at
$\omega_0$.

{ 
At this stage, the synthesis problem can be carried out in the normalized low-pass domain $\hat s$, by designing a \textit{low-pass} prototype. For the scope of the Letter, we consider the Cauer ladder realization shown in Fig. \ref{fig:lpproto}(b). 
}

Let $(g_n)_{n=1}^N$ denote its normalized series inductance ($n$ odd) and shunt capacitances ($n$ even). The normalized terminating resistance is $g_{N+1} = r_m$, ensuring the prescribed material resistance at $\omega_0$, since the series inductors are short circuits and the shunt capacitors are open circuits at $\hat{s}=0$.
The associated reflection coefficient is denoted by
$\hat{\Gamma}
=\hat{\Gamma}(\hat{s};g_1,\dots,g_{N+1})$.
Because the ladder is lossless, its power transmission coefficient equals the normalized absorption response, $\hat P_\abs/\hat P_{\rm av}
=1-|\hat{\Gamma}|^2$.

Standard prototypes compatible with the prescribed
termination can therefore be used to shape the
absorption spectrum. Butterworth and Chebyshev
designs prescribe the transmission magnitude,
whereas Bessel designs prioritize a maximally flat
group delay
\cite{pozar_microwave_2012,butterworth_theory_1930,
thomson_delay_1949, matthaei_synthesis_1956}.

From this normalized response, the ladder parameters
$(g_n)_{n=1}^{N+1}$ are obtained using standard
matching-network synthesis techniques
\cite{pozar_microwave_2012,su_analog_1996}.
For many standard prototypes, including Butterworth,
Chebyshev, and Bessel responses, these parameters
are available in closed form or tabulated in
low-pass prototype tables
\cite{pozar_microwave_2012}.
Once the ladder parameters are known, they determine the reflection
coefficient $\hat\Gamma$, and hence the
{ normalized driving-point impedance
$\hat Z_{\rm in}$}.
Finally, applying the
{ inverse frequency transformation}
\eqref{eq:Lp2Bp} to this prototype yields
a physical band-pass network in the $s$-domain.
{ The series inductors and shunt capacitors
of the low-pass ladder map onto
series and parallel LC resonators, respectively,
all tuned to $\omega_0$ \cite{SuppMat}.} { Having designed $Z_\chi$,} the material susceptibility follows via
$\chi^{-1}=jka\,Z_\chi/Z_0$
\cite{forestiere_first-principles_2024}.
Other details on the synthesis techniques are
provided in the Supplemental Material
\cite{SuppMat}.


\begin{figure}
    \centering
    \includegraphics[width=1\linewidth]{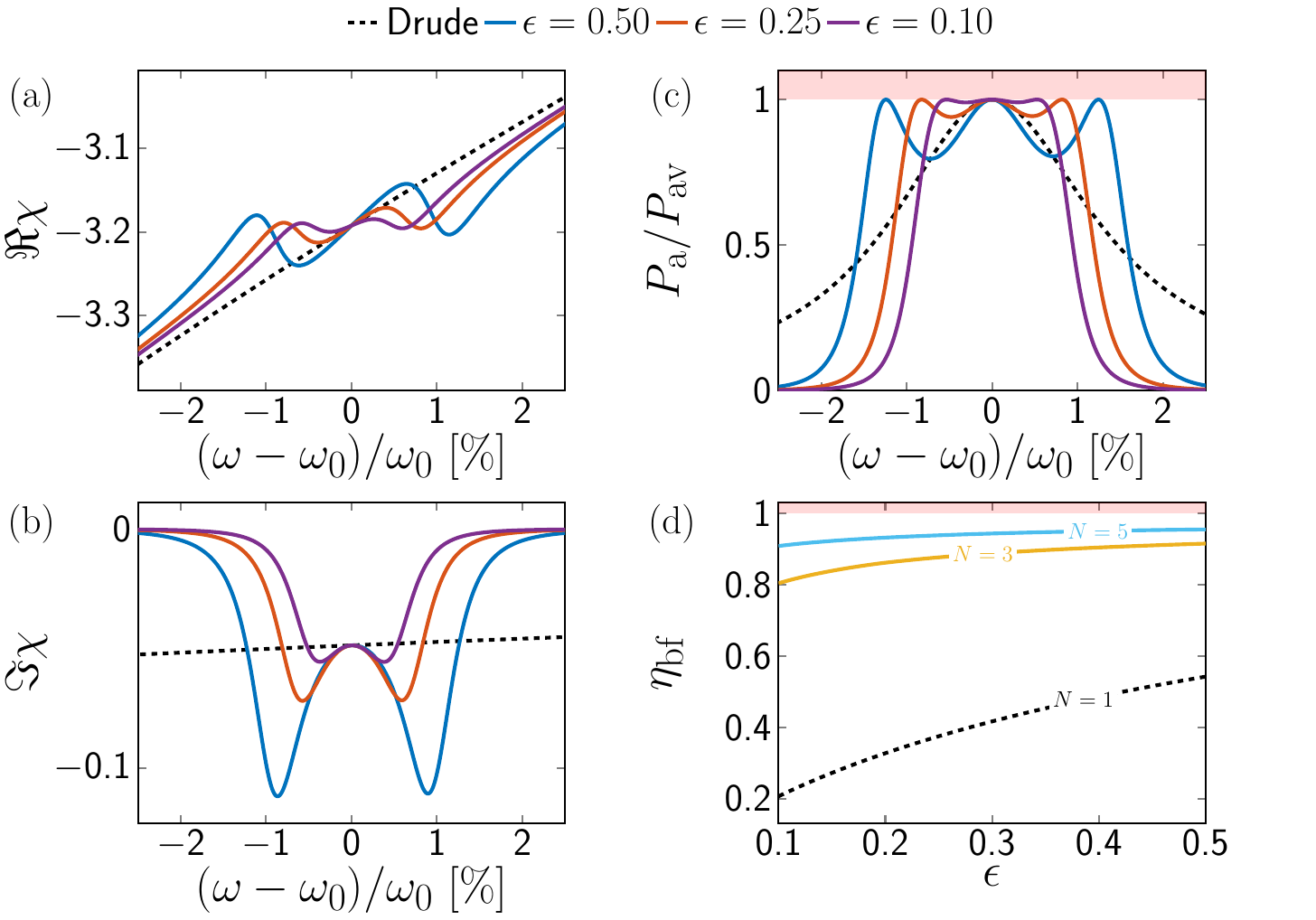}
   \caption{Chebyshev synthesis approaching the Bode--Fano integral bound. (a) Real and (b) imaginary parts of the susceptibility for third-order
passive causal realizations with ripple parameter
\(\epsilon=0.50,0.25,0.10\).  The dashed black curve is the matched
single-pole Drude reference, \(N=1\).  (c) Normalized absorbed power
\(P_{\abs}/P_{\rm av}\), with both powers
evaluated using full-wave Mie theory.
The shaded region marks \(P_{\abs}>P_{\rm av}\), which is forbidden by passivity within the single-mode approximation.
Reducing \(\epsilon\) improves the uniformity of the near-conjugate match
inside the band but narrows the achievable bandwidth.  (d) Normalized
Bode--Fano integral \(\eta_{\rm BF}\) versus ripple parameter. The shaded region marks \(\eta_{\rm BF}>1\), which is forbidden by the Bode--Fano limit.  The higher-order Chebyshev realizations approach saturation of the in-band integral budget more than the matched Drude response.}
    \label{fig:chebyshev}
\end{figure}

\paragraph{Results---}

Our formulation enables the syntesis of material dispersion to approach the Bode-Fano integral bounds for a fixed scatterer geometry. We consider a sphere of radius \(a\). { This geometry is especially relevant { for the electric dipole} because, among all bodies contained in the same circumscribing sphere, it attains the minimum value of the quality factor
$Q_{\min}$, thereby saturating
the geometric inequality underlying
Eq.~\eqref{eq:bode-fano-tradeoff}.
} 

First, we consider the minimal realization, \(N=1\), satisfying the matching condition \(R_m=R_s\).  The material impedance is then
a series \(RL\) circuit, $Z_\chi = R_m + j \omega L_m$, corresponding to a Drude susceptibility $\chi^{-1}=-{\omega(\omega-j\gamma)}/{\omega_p^2}$ \cite{kelly_optical_2003}, with plasma frequency $\omega_p = 1/\sqrt{L_m \varepsilon_0a}$ and damping rate $\gamma = R_m/L_m$. For a prescribed $\omega_p$, the inductance $L_m$
is fixed, the zero-reactance condition determines
$\omega_0$, and resistance matching sets
$\gamma=\mathcal R_{\ed}(\omega_0)/L_m$.  In the calculations below, we choose \(\omega_p\sqrt{\mu_0\varepsilon_0}\,a=0.5\), which gives \(\gamma/\omega_p=8.5443\times10^{-3}\).  The corresponding Drude susceptibility is shown by the dashed curves in Fig.~\ref{fig:chebyshev}(a,b).

We next synthesize higher-order dispersion relations that comply with passivity and causality constraints, that approach conjugate matching over a finite bandwidth. The target absorbed-power response is chosen as a Type-I Chebyshev form, that is $\hat{P}_\abs/\hat P_{\rm av} = 1/\left[1+\epsilon^2
T_N^2\!\left(\hat\omega/\hat\omega_c\right)\right]$, where \(T_N\) is the Chebyshev polynomial of the first kind  and $\epsilon>0$ controls the in-band ripple.  Within $|\hat\omega|\leq\hat\omega_c$, the response oscillates between $1-\delta$ and unity, with $\delta=\epsilon^2/(1+\epsilon^2)$.
This equiripple design therefore produces equal
maxima of the power mismatch across the passband.

Figure~\ref{fig:chebyshev}(a,b) shows the susceptibilities for third-order realizations for three
ripple levels, namely $\epsilon = 0.5$ (blue line), $\epsilon = 0.25$ (red), and $\epsilon = 0.1$ (purple). For these designs, Fig. \ref{fig:chebyshev}(c) shows the normalized total absorbed power $P_{\abs}/P_{\rm av}$ under a plane wave illumination. Both \(P_{\abs}\) and \(P_{\mathrm{av}}\) are computed from full-wave Mie theory~\cite{mie_beitrage_1908} including multipolar order beyond the dipole: $P_{\abs}$ is evaluated for the synthesized material, while \(P_{\mathrm{av}}\) is the
reference, obtained by imposing the ideal conjugate-matching condition \(Z_{\chi}=\zeta_{\ed}^{*}\) at each frequency.   For these designs, reducing $\epsilon$ improves
in-band matching while narrowing the absorption band. Indeed, smaller ripple implies smaller \(|\Gamma|\) throughout the band and therefore a larger logarithmic term \(\ln(1/|\Gamma|)\) in the Bode--Fano constraint \eqref{eq:bode-fano}. Consequently, the bandwidth must shrink as the response approaches ideal conjugate matching.  To quantify how closely the synthesized responses approach the Bode--Fano limit, we define the normalized in-band Bode--Fano integral
\begin{equation*}
   \eta_{\rm BF}
:=
{\frac{3}{2 \pi} \left(\frac{a}{c_0}\right)^{-3}}
\int_{\omega_1}^{\omega_2}
\ln\frac{1}{|\Gamma(j\omega)|}\,
\frac{\mathrm{d}\omega}{\omega^{ 4}}.
\end{equation*}
Here, $\Gamma$ is evaluated using the full-wave electric-dipole
radiation impedance, and the band edges $\omega_1$ and $\omega_2$
are determined from the normalized dipolar absorption spectrum
$P_a/P_{\mathrm{av}}=1-|\Gamma|^2$ at the threshold
$1/(1+\epsilon^2)$.

Figure~\ref{fig:chebyshev}(d)
reports \(\eta_{\rm BF}\) as a function of the ripple parameter \(\epsilon\)
for three synthesis orders, namely $N=1$ (black dashed line), $N=3$ (yellow), and $N=5$ (cyan).  Figure~\ref{fig:chebyshev}(d) shows that the Chebyshev realizations use the
available Bode--Fano integral far more efficiently than the Drude response; for \(N=5\), \(\eta_{\rm BF}\) remains in the range $[0.91,0.96]$ when \(\epsilon \in [0.1,0.5]\), indicating
proximity to the small-particle Bode--Fano limit.

\paragraph{Conclusions---}
{
We have derived a Bode--Fano-type absorption--bandwidth bound for electrically small scatterers dominated by a single electric-dipole mode by connecting electromagnetic scattering with classical matching-network theory. The same framework provides a constructive route to synthesizing material dispersion that approaches the Bode-Fano integral bounds for a prescribed scatterer size.
}
Building on the equivalent-circuit description of Ref.~\cite{forestiere_first-principles_2024},  the formulation separates radiation and material contributions through
their respective impedances and expresses normalized absorption in terms of a reflection coefficient.  The derivation assumes passive, causal, linear, time-invariant material responses with rational susceptibility and uses a rational small-particle approximation to the radiation impedance. The resulting bound quantifies the trade-off between absorption level and bandwidth and connects it to quasistatic quality-factor bounds for electric-type resonators.

Beyond establishing fundamental bounds, our formulation enables a rational synthesis route for material dispersion that, while complying with causality and passivity, approaches Bode-Fano integral bounds. Darlington synthesis first expresses the physically realizable material impedance as a lossless two-port network terminated by a resistor. A subsequent band-pass/low-pass transformation maps the broadband absorption design to the optimization of a normalized low-pass prototype. The resulting higher-order susceptibilities { provide passive and causal target material responses that } approach the Bode--Fano bounds and explicitly display the fundamental trade-off between high-absorption level and bandwidth. Possible routes to approximating these responses over a finite band include quantum-engineered resonant systems \cite{lee_giant_2014,mann_ultrafast_2021} and multiresonant composites or metamaterials \cite{Han23,li_lorentz_2021,Alu11}.  We have accurately validated the synthesized responses against full-wave Mie calculations.

We have accurately validated the synthesized responses against full-wave Mie calculations. 

These results provide a basis for designing material dispersion for broadband light harvesting and controlled dissipation. The framework also extends to other isolated modes when their radiation impedances satisfy the required analytic and asymptotic conditions, as demonstrated by the magnetic-dipole bound derived in the Supplemental Material \cite{SuppMat}.

More broadly, the results quantify the restrictions
that passivity and causality impose on broadband
absorption within the stated assumptions.
Active, time-varying, or nonlinear media may enable
responses beyond these limits by relaxing the
underlying assumptions
\cite{li_beyond_2019,hayran_beyond_2024,
shlivinski_beyond_2018}.

\paragraph{Acknowledgments---} E. C. and C. F. acknowledge support from the Italian Ministry of University and Research (MUR) through the PRIN 2022 project (Grant No. 2022Y53F3X, “Inverse Design of High-Performance Large-Scale Metalenses”), and from the Università degli
Studi di Napoli Federico II through its International Agreements Support Grant (Board Resolution No. 55, February 12, 2025). A. A. acknowledges support from
the Simons Foundation.

\nocite{forestiere_volume_2018,forestiere_material-independent_2016,mayergoyz_electrostatic_2005,forestiere_resonance_2020}


\providecommand{\href}[2]{#2}

\clearpage
\onecolumngrid
\setcounter{section}{0}
\setcounter{subsection}{0}
\setcounter{subsubsection}{0}
\setcounter{equation}{0}
\setcounter{figure}{0}
\setcounter{table}{0}
\renewcommand{\theequation}{S\arabic{equation}}
\renewcommand{\thefigure}{S\arabic{figure}}
\renewcommand{\thetable}{S\arabic{table}}
\renewcommand{\thesubsection}{\Roman{section}.\arabic{subsection}}

\begin{center}
  {\large\bfseries Supplemental Material for:\par}
  \vspace{0.35em}
  {\large\bfseries Bode--Fano Limits to Broadband Absorption by Small Particles\par}
  \vspace{1em}
  {\normalsize Emanuele Corsaro$^{1,2}$, Andrea Al\`u$^{2,3}$, and Carlo Forestiere$^{1}$\par}
  \vspace{0.6em}
  {\small $^1$Department of Electrical Engineering and Information Technology,\\
  Universit\`a degli Studi di Napoli Federico II, via Claudio 21, Napoli, 80125, Italy\par}
  {\small $^2$Photonics Initiative, Advanced Science Research Center,\\
  City University of New York, New York, New York 10031, USA\par}
  {\small $^3$Physics Program, Graduate Center of the City University of New York,\\
  New York, New York 10016, USA\par}
\end{center}

\tableofcontents
\clearpage

\section{Circuit model of the electromagnetic scattering}
\label{sec:CircuitModel}
Consider a linear, homogeneous, isotropic, and time-dispersive object with susceptibility $\chi(\omega)$ embedded in vacuum.
Let $\Omega \subset \mathbb{R}^3$ be the bounded domain occupied by the object and $\partial \Omega$ its boundary.  Let $a$ be the radius of the smallest ball enclosing the object.  The outward-pointing unit vector on $\partial \Omega$ is denoted by $\hat{\mathbf{n}}$. The object is excited by an incident time-harmonic electric field $\mathbf{E}_{\inc}$, with real angular frequency (a time-harmonic dependence $e^{j\omega t}$ is assumed). 

The full-wave scattering problem can be formulated considering as unknown the polarization current density field $\mathbf{J}$ induced in the object by the incident field $\mathbf{E}_{\inc}$. In $\Omega$, the constitutive relation links the field ${\bf J}$ to the total electric field  ${\mathbf{E}} = \vec{E}_\inc + \vec{E}_\sca$, which is the sum of the scattered field ${\mathbf{E}}_{\sca}$ and the incident field ${\mathbf{E}}_{\inc} $, by the relation
\begin{equation}
a \, Z_\chi {\bf J}  = {\bf E}_{\sca} +{\bf E}_{\inc}  \qquad \forall \mathbf{r} \in \Omega,
 \label{seq:constrel}
\end{equation}
where  $Z_\chi = Z_\chi(j\omega)$ is the \textit{material impedance} defined as \cite{forestiere_first-principles_2024}
 \begin{equation}
    Z_\chi := \frac{ Z_0}{jk a\, \chi}.
    \label{seq:Zmat}
\end{equation}
Here, $Z_0 = \sqrt{\mu_0/\varepsilon_0}$ is the vacuum characteristic impedance, $k = \omega \sqrt{\varepsilon_0 \mu_0}$, $\varepsilon_0$, $\mu_0$ are the vacuum permittivity and permeability, respectively. The real and imaginary parts of $Z_\chi(j\omega)$  are called \textit{material resistance} ($R_\chi$) and \textit{material reactance} ($X_\chi$), respectively. Note that both the vector fields ${\bf E}_{\sca}$ and ${\bf J}$ are divergence-free within $\Omega$ due to the homogeneity and isotropy of the material.

The scattered electric field $\vec{E}_\sca$ is related to the polarization current $\vec{J}$ as $\vec{E}_\sca = - a \mathcal{L}_\Omega\{\vec{J}\}$, where $\mathcal{L}_\Omega$ is the integral operator defined as
\begin{equation}
  \mathcal{L}_\Omega\{ \mathbf{J} \}(\mathbf{r}) := 
    \frac{Z_0}{j k a} \nabla_\mathbf{r} \oint_{\partial \Omega} g(\mathbf{r} - \mathbf{r}') \, \mathbf{J}(\mathbf{r}') \cdot \hat{\mathbf{n}}(\mathbf{r}') \, \mathrm{d}S' \\
    + j k Z_0 \frac{1}{a} \int_{\Omega} g(\mathbf{r} - \mathbf{r}') \, \mathbf{J}(\mathbf{r}') \, \mathrm{d}\Omega',
   \label{seq:Loperator}
\end{equation}
where $g$ is the homogeneous space Green’s function, i.e.
\begin{equation*}
    g(\vec{r} - \vec{r}') = \frac{e^{-j k |\vec{r}-\vec{r}'|}}{4\pi  |\vec{r}-\vec{r}'|}.
\end{equation*}
We expand the induced current density $\vec{J}$, solution of the scattering problem \eqref{seq:constrel}, in terms of the polarization current density modes $\left\{ \mathbf{j}_h \right\}_{h \in \mathbb{N}}$ \cite{forestiere_first-principles_2024}:
\begin{equation}
\vec{J}  = \frac{1}{a^2} \sum_{h = 1}^{\infty} I_h(\omega) \, \vec{j}_h(\vec{r}, \omega).
\label{seq:MIMexpansion}
\end{equation}
 The current density modes $ \mathbf{j}_h $ are the eigenmodes of  $\mathcal{L}_\Omega$ in $\Omega$  \cite{forestiere_first-principles_2024,forestiere_volume_2018,forestiere_material-independent_2016}:
\begin{equation}
    \mathcal{L}_\Omega \mathbf{j}_h = \zeta_{\Omega,h} \, \mathbf{j}_h,
    \label{seq:EigProb}
\end{equation}
where $\zeta_{\Omega,h} = \mathcal{R}_{\Omega,h} + j \mathcal{X}_{\Omega,h}$ is the \textit{radiation impedance} of the current eigenmode $\mathbf{j}_h$, $\mathcal{R}_{\Omega,h}$ is the \textit{radiation resistance}, and $\mathcal{X}_{\Omega,h}$ is the \textit{radiation reactance}. 
The current modes $\vec{j}_h$ are not orthogonal with respect to the standard inner product over $\Omega$, that is
\begin{equation*}
    \langle \vec{u},\vec{v}\rangle := \frac{1}{a^3} \int_\Omega \vec{u}^*(\vec{r}) \cdot \vec{v}(\vec{r}) \dV,
\end{equation*}
but they are bi-orthogonal, namely $\langle\vec{j}_k^*, \vec{j}_h\rangle = 0$ for $h \neq k$. Without loss of generality, we impose the normalization condition $\|\vec{j}_h\|^2 = 1$, where $\|\vec{j}_h\|^2:= \langle \vec{j}_h, \vec{j}_h \rangle$.

The coefficients $I_h$, which have the dimension of an electric current intensity, are given by
\begin{equation}
    I_h  = \frac{\mathcal{E}_h }{ Z_\chi  + \zeta_{\ed,h}  },
    \label{seq:CurrentI}
\end{equation}
where the equivalent voltage source   $\mathcal{E}_h$ is
\begin{equation}
    \mathcal{E}_h = a\frac{\langle   \vec{j}_h^*, {{\bf E}_{\inc}} \rangle}{\langle   \vec{j}_h^*, \vec{j}_h \rangle}.
    \label{seq:volt_filter}
\end{equation}
Finally, the polarization current modes $\{\vec{j}_h\}_{h=1}^{\infty}$ are also used to express the incident, the scattered, and the total electric fields in $\Omega$ as
\begin{equation}
    \vec{E}_\inc = \frac{1}{a} \sum_{h=1}^\infty \mathcal{E}_h (\omega) \vec{j}_h(\vec{r},\omega), \qquad  \vec{E}_\sca = -\frac{1}{a} \sum_{h=1}^\infty  V_{\ed,h} (\omega) \vec{j}_h(\vec{r},\omega), \qquad   \vec{E} = \frac{1}{a} \sum_{h=1}^\infty  V_{\chi,h} (\omega) \vec{j}_h(\vec{r},\omega),
    \label{seq:E_expansion}
\end{equation}
with $V_{\ed,h} := \zeta_{\ed,h}  I_h$, and $V_{\chi,h} := Z_{\chi}  I_h$.

In the above circuit representation, the dependencies on the excitation condition, the geometry, and the material of the object are disentangled: the geometry is accounted for by the radiation impedance $\zeta_{\Omega,h}$, which does not depend on the material of the object, but only on its shape and the frequency. The material is accounted for by the material impedance $Z_\chi$.
The excitation condition is accounted for by the voltage source $\mathcal{E}_h$.

\subsection{Scattering, Absorption and Extinction efficiencies in the dipolar regime}

The time-averaged power scattered, absorbed, and extinguished by the object are
\begin{equation}
    P_{\sca} = - \frac{a^3}{2}  \Re \langle \vec{J}, \vec{E}_{\sca} \rangle, \qquad     P_{\abs} =  \frac{a^3}{2}  \Re \langle \vec{J}, \vec{E} \rangle, \qquad P_{\ext} = \frac{a^3}{2}  \Re \langle \vec{J}, \vec{E}_{\inc} \rangle,
    \label{seq:opt_Powers0}
\end{equation}
respectively.  The ratio of these powers to  $P_\inc = \pi a^2|\vec{E}_{\inc}|^2/(2Z_0)$, returns the scattering ($Q_{\sca}$), absorption ($ Q_{\abs}$), and extinction ($Q_{\ext}$) efficiencies \cite{bohren_absorption_1998}.

When scattering is dominated by a single mode, the expansions in Eqs. \eqref{seq:MIMexpansion} and \eqref{seq:E_expansion} reduce to the single term. Consequently, Eq. \eqref{seq:opt_Powers0} becomes
\begin{equation}
    P_{\sca} =  \frac{1}{2} \mathcal{R}_{\ed} |I|^2, \qquad     P_{\abs} =   \frac{1}{2} R_{\chi} |I|^2, \qquad P_{\ext} = \frac{1}{2} \Re \{I^* \mathcal{E}\},
    \label{seq:opt_Powers}
\end{equation}
where $I = I_h$, $\mathcal{E} = \mathcal{E}_h$, $Y = (Z_\chi+ \zeta_{\Omega,h})^{-1}$. Since $I = Y\mathcal{E}$, with $Y = G + j B$ one obtains
\begin{equation}
    P_{\sca} =  \frac{1}{2} \mathcal{R}_{\ed} |Y|^2 |\mathcal{E}|^2 , \qquad     P_{\abs} =   \frac{1}{2} R_{\chi} |Y|^2|\mathcal{E}|^2, \qquad P_{\ext} = \frac{1}{2} G|\mathcal{E}|^2.
    \label{seq:opt_Powers2}
\end{equation}

\subsection{Positive-real property of the radiation impedance}

We first establish the positive-real character of the radiation
impedance $\zeta_{\Omega,h}(s)$. Recall that a function $f(s)$ is
positive real if it is analytic in the open right-half plane
$\mathbb C_+:=\{s\in\mathbb C:\Re s>0\}$, satisfies
$\Re f(s)\geq0$ for $s\in\mathbb C_+$, and is real for $s>0$.

Consider a polarization current supported in $\Omega$ and proportional
to the $h$th current-density mode,
\begin{equation}
    \mathbf J_h = J_h \mathbf j_h .
\end{equation}
Let $\mathbf E_h^{\sca}$ and $\mathbf H_h^{\sca}$ denote the
corresponding radiated fields. In the Laplace domain they satisfy
\begin{subequations}
\begin{align}
    \nabla\times{\mathbf E}_h^{\sca}
    &=
    -s\mu_0{\mathbf H}_h^{\sca},
    \label{eq:modal_maxwell_E}
    \\
    \nabla\times{\mathbf H}_h^{\sca}
    &=
    {\mathbf J}_h
    +s\varepsilon_0{\mathbf E}_h^{\sca}.
    \label{eq:modal_maxwell_H}
\end{align}
\end{subequations}
Defining the complex Poynting vector
\begin{equation}
    {\vec{S}}_h = \frac{1}{2}({\vec{E}}_h^\sca \times {\vec{H}}_h^{\sca*}),
\end{equation}
the Poynting theorem takes the form
\begin{equation}
    -\int_\Omega
    {\mathbf J}_h^*\cdot
    {\mathbf E}_h^{\sca}
    \dV
    = 2\oint_{\mathbb{S}_\infty} {\vec S}_h \cdot \vers{n} \dS+
    \int_{\mathbb R^3}
    \left(
    \varepsilon_0s^*|{\mathbf E}_h^{\sca}|^2
    +
    \mu_0s|{\mathbf H}_h^{\sca}|^2
    \right)
    \dV.
    \label{eq:laplace_poynting_infinity}
\end{equation}
Inside the scatterer,
\begin{equation}
    {\mathbf E}_h^{\sca}
    =
    -a\zeta_{\Omega,h}{\mathbf J}_h.
    \label{eq:E_mode_zeta}
\end{equation}
For $s\in\mathbb C_+$ the radiated fields decay exponentially at spatial infinity. Consequently, in $\mathbb C_+$, the surface integral in Eq. \eqref{eq:laplace_poynting_infinity} vanishes at spatial infinity.  Thus, writing
$s=\sigma+ j\omega$, Eq.~\eqref{eq:laplace_poynting_infinity} for $\sigma>0$ becomes
\begin{equation}
\begin{split}
  a^4  \|{\vec{J}}_h\|^2\zeta_{\Omega,h}
    ={}&
    \sigma
    \int_{\mathbb R^3}
    \left(
    \varepsilon_0|{\mathbf E}_h^{\sca}|^2
    +
    \mu_0|{\mathbf H}_h^{\sca}|^2
    \right)
    \dV
    +
    j\omega
    \int_{\mathbb R^3}
    \left(
    \mu_0|{\mathbf H}_h^{\sca}|^2
    -
    \varepsilon_0|{\mathbf E}_h^{\sca}|^2
    \right)
    \dV.
\end{split}
\label{eq:zeta_energy_identity}
\end{equation}
Let
\begin{equation}
    \vec{e}_h^\sca = \frac{\vec{E}_h^s}{a^2\|\vec{J}_h\|}, \qquad \vec{h}_h^\sca = \frac{\vec{H}_h^\sca}{a^2\|\vec{J}_h\|};
\end{equation}
the real and imaginary of $\zeta_{\Omega,h}$ in the open right half plane are
\begin{align}
   \Re\{\zeta_{\Omega,h}\}
   &=
    \sigma
    \int_{\mathbb R^3}
    \left(
    \varepsilon_0|{\mathbf e}_h^{\sca}|^2
    +
    \mu_0|{\mathbf h}_h^{\sca}|^2
    \right)
    \dV.
    \label{eq:zeta_energy_real}
\\
   \Im\{\zeta_{\Omega,h}\}
    &=
    \omega
    \int_{\mathbb R^3}
    \left(
    \mu_0|{\mathbf h}_h^{\sca}|^2
    -
    \varepsilon_0|{\mathbf e}_h^{\sca}|^2
    \right)
    \dV.
    \label{eq:zeta_energy_imag}
\end{align}
The right-hand side of Eq. \eqref{eq:zeta_energy_real} is nonnegative
\begin{equation}
    \Re\{\zeta_{\Omega,h}(s)\}\geq0,
    \qquad
    s\in\mathbb C_+.
    \label{eq:zeta_positive}
\end{equation}
Morover, from Eq. \eqref{eq:zeta_energy_imag} it follows that $\Im\{\zeta_{\Omega,h}\} = 0$ for $\omega =0$, i.e., $\zeta_{\Omega,h}$ is real on the real axis.  Thus, the radiation impedance
$\zeta_{\Omega,h}(s)$ is a positive-real function.


\subsection{Asymptotic expansions and  small-particle limit of the radiation impedance}

\begin{figure}
    \centering
    \includegraphics[width=0.75\linewidth]{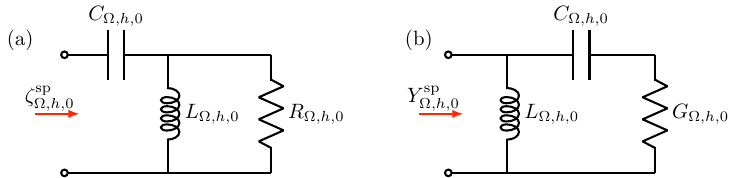}
    \caption{Passive circuit representation of the small-particle
    approximation  of the
    radiation impedance { for plasmonic (a) and dielectric modes (b).} }
    \label{fig:Zed_laurent}
\end{figure}

{
Based on their low-frequency behaviour, the polarization current modes can be classified into two families, called \emph{plasmonic} and \emph{dielectric} modes \cite{forestiere_first-principles_2024}.

Plasmonic modes, such as the electric dipole, electric quadrupole, electric octupole, and so on, have ohmic-capacitive radiation impedance. Their expansion about $s=0$ and $s=\infty$ read
}
\begin{subequations}
\label{seq:Laurent_Zed}
\begin{align}
    \zeta_{\ed,h}(s)
    &=
    \frac{1}{sC_{\ed,h,0}}
    + sL_{\ed,h,0}
    - s^2r_{\ed,h,0}
    + O(s^3),
    && s\to 0,
    \label{seq:Lau0_ed}\\
    \zeta_{\ed,h}(s)
    &=
    \frac{1}{sC_{\infty}}
    + O(s^{-2}),
    && s\to\infty,
    \label{seq:Lauinf_ed}
\end{align}
\end{subequations}
where $C_{\ed,h,0}>0$, $L_{\ed,h,0}>0$, $r_{\ed,h,0}\geq0$, $C_{\infty} = \varepsilon_0 a$.
The leading capacitive term in Eq.~\eqref{seq:Lau0_ed} describes
the electrostatic response, while $sL_{\ed,h,0}$ and $-s^2r_{\ed,h,0}$
give the leading inductive and radiative corrections, respectively.
In particular, on the imaginary-frequency axis $s=j\omega$,
the latter contributes the positive radiation resistance
$R_{\Omega,h}(\omega) =\omega^2r_{\ed,0} + O(\omega^4)$.

{
Dielectric modes, such as the magnetic dipole, magnetic quadrupole, and the toroidal electric dipole, have ohmic-inductive radiation impedance. The expansion of their radiation admittance $Y_{\Omega,h}=\zeta_{\Omega,h}^{-1} $ obout $s=0$ and $s=\infty$ reads
\begin{subequations}
\label{seq:Laurent_Zed}
\begin{align}
    Y_{\Omega,h}(s)
    &=
     \frac{1}{sL_{\ed,h,0}} + sC_{\Omega,h,0}
    - s^2g_{\ed,h,0}
    + O(s^3),
    && s\to 0,
    \label{seq:Lau0_md}\\
    Y_{\Omega,h}(s)
    &=
   sC_{\ed,\infty}
    + O(s^{-2}).
    && s\to\infty,
    \label{seq:Lauinf_md}
\end{align}
\end{subequations}
where $L_{\ed,h,0}>0$, and $g_{\ed,h,0}\geq 0$.
The leading iductive term in Eq.~\eqref{seq:Lau0_md} describes
the magnetostatic response, while $-s^2g_{\ed,h,0}$
gives the leading radiative correction.
In particular, on the imaginary-frequency axis $s=j\omega$,
the latter contributes the positive radiation conductance
$G_{\Omega,h}(\omega) = \omega^2g_{\ed,0} + O(\omega^4)$.
}

A passive rational approximation that reproduces
Eqs.~\eqref{seq:Lau0_ed} { and \eqref{seq:Lau0_md}} through order $s^2$ are
\begin{equation}
    \zeta_{\ed,h}^{\mathrm{sp}}(s)
    =
    \frac{1}{sC_{\ed,h,0}}
    +
    \frac{sL_{\ed,h,0}R_{\ed,h,0}}
         {R_{\ed,h,0}+sL_{\ed,h,0}},
    \qquad
    R_{\ed,h,0}:=\frac{L_{\ed,h,0}^2}{r_{\ed,h,0}}>0,
    \label{eq:zeta_qs}
\end{equation}
{ and
\begin{equation}
    Y_{\Omega,h}^{\mathrm{sp}}(s)
    =
    \frac{1}{sL_{\ed,h,0}}
    +
    \frac{sC_{\ed,h,0}G_{\ed,h,0}}
         {G_{\ed,h,0}+sC_{\ed,h,0}},
    \qquad
    G_{\ed,h,0}:=\frac{C_{\ed,h,0}^2}{g_{\ed,h,0}}>0,
    \label{eq:zeta_qs_diel}
\end{equation}
respectively.}
Eq. \eqref{eq:zeta_qs} corresponds to the impedance of a capacitor $C_{\ed,h,0}$ in series
with the parallel combination $L_{\ed,h,0}\parallel R_{\ed,h,0}$,
as shown in Fig.~\ref{fig:Zed_laurent}(a); { whereas Eq. \eqref{eq:zeta_qs_diel} corresponds to the impedance of an inductor $L_{\ed,h,0}$ in parallel
with the series combination $C_{\ed,h,0}  R_{\ed,0}$,
as shown in Fig.~\ref{fig:Zed_laurent}(b).}
We refer to Eqs.~\eqref{eq:zeta_qs}-\eqref{eq:zeta_qs_diel} as the
\emph{small-particle approximation} of the radiation impedance.
Its matching to the exact impedance is local to $s=0$:
it does not reproduce the high-frequency behavior in
Eqs.~\eqref{seq:Lauinf_ed}-\eqref{seq:Lauinf_md}.

For a general geometry, the coefficients
$C_{\ed,h,0}$, $L_{\ed,h,0}$, and $r_{\ed,h,0}$ can be evaluated
using the electroquasistatic mode and the low-frequency expansion
of the radiation operator, without computing corrections to
the mode itself. To establish this result, introduce
\begin{equation}
    \mathcal T_\Omega(s)
    :=
    s\,\mathcal L_\Omega(s).
\end{equation}
{ Thus, the eigenvalue problem \eqref{seq:EigProb} becomes
\begin{equation}
    \mathcal T_\Omega(s)\mathbf j_{h}(s)
    =
    \lambda_{h}(s)\mathbf j_{h}(s),
    \qquad
    \lambda_{h}(s)
    =
    s\zeta_{\Omega,h}(s).
    \label{seq:A_eig}
\end{equation}
}
Expanding the free-space Green function about $s=0$ gives
\begin{equation}
    \mathcal T_\Omega(s)
    =
    \mathcal T_0
    +s^2\mathcal T_2
    +s^3\mathcal T_3
    +O(s^4),
    \label{seq:A_perturbation}
\end{equation}
where $\mathcal T_0$ is the Hermitian electrostatic operator
{
\begin{equation}
 \mathcal{T}_0 \vec{J} :=    \frac{1}{4\pi \varepsilon_0 a} \nabla_\vec{r} \oint_{\partial \Omega} \frac{\vers{n}(\vec{r}')\cdot\vec{J}(\vec{r}')}{|\vec{r}-\vec{r}'|}\dS'.
 \label{seq:T0def}
\end{equation}
}
The term linear in $s$ in the Green-function expansion is
spatially constant and is therefore annihilated by the gradient
in the scalar-potential contribution.
The vector-potential contribution starts at order $s^2$, { through 
\begin{equation}
    \mathcal{T}_2  \vec{J}  = \frac{1}{8\pi \varepsilon_0 a c_0^2} \nabla_{\vec{r}}\oint_{\partial \Omega} {|\vec{r}-\vec{r}'|}\vers{n}(\vec{r}')\cdot\vec{J}(\vec{r}') \dS' + \frac{\mu_0}{4\pi a}\int_{\Omega}\frac{\vec{J}(\vec{r}')}{|\vec{r}-\vec{r}'|}\dV',
\end{equation}
and
\begin{equation}
    \mathcal{T}_3\vec{J} = -\frac{1}{6\pi \varepsilon_0 a c_0^3}\int_{\Omega} \vec{J}(\vec{r}') \dV'.
    \label{seq:T3def}
\end{equation}
}

{

We now distinguish the perturbative expansions associated
with plasmonic and dielectric modes. In the space of
divergence-free current densities, the electrostatic operator
$\mathcal T_0$ is self-adjoint and has a nontrivial kernel.
The nonzero eigenvalues of $\mathcal T_0$ generate the
plasmonic branches, whereas its kernel contains the
magnetoquasistatic current modes associated with the
dielectric branches.

For simplicity, we consider regular modal branches
associated with isolated quasistatic eigenvalues.
When degeneracies occur, the corresponding perturbations
are understood to be resolved within the degenerate
eigenspaces.

\paragraph*{Plasmonic modes.}

Let $\{\mathbf j_h^\parallel\}$ denote the orthonormal
electroquasistatic current modes associated with the
nonzero eigenvalues of $\mathcal T_0$, namely
\begin{equation}
    \mathcal T_0\mathbf j_h^\parallel
    =
    \lambda_h^{(0)}\mathbf j_h^\parallel,
    \qquad
    \lambda_h^{(0)}>0.
    \label{seq:EQS_eig}
\end{equation}
Their properties are discussed in
Ref.~\cite{mayergoyz_electrostatic_2005}.
For the perturbative calculation
(see Ref.~\cite{forestiere_resonance_2020}),
we adopt the intermediate normalization
$\langle\mathbf j_h^\parallel,\mathbf j_h(s)\rangle=1$.

Analytic perturbation theory gives
\begin{subequations}
\begin{align}
    \lambda_h(s)
    &=
    \lambda_h^{(0)}
    +s^2\lambda_h^{(2)}
    +s^3\lambda_h^{(3)}
    +O(s^4),
    \\
    \mathbf j_h(s)
    &=
    \mathbf j_h^\parallel+O(s^2).
\end{align}
\label{seq:EQS_pert}
\end{subequations}
Substituting these expansions into
Eq.~\eqref{seq:A_eig} and projecting onto
$\mathbf j_h^\parallel$, the self-adjointness
of $\mathcal T_0$ yields
\begin{equation}
    \lambda_h^{(2)}
    =
    \left\langle
    \mathbf j_h^\parallel,
    \mathcal T_2\mathbf j_h^\parallel
    \right\rangle,
\qquad     \lambda_h^{(3)}
    =
    \left\langle
    \mathbf j_h^\parallel,
    \mathcal T_3\mathbf j_h^\parallel
    \right\rangle.
\label{seq:EQS_lambda23}
\end{equation}
Since $\zeta_{\ed,h}(s)=\lambda_h(s)/s$,
comparison with Eq.~\eqref{seq:Lau0_ed} gives
\begin{equation}
    C_{\ed,h,0}^{-1}
    =\lambda_h^{(0)},\qquad
    L_{\ed,h,0}
    =\lambda_h^{(2)},\qquad
    r_{\ed,h,0}
    =-\lambda_h^{(3)}.
\label{seq:EQS_parameters}
\end{equation}
Thus, the low-frequency radiation impedance
through order $s^2$ depends only on the
unperturbed electroquasistatic mode and the
operators $\mathcal T_0$, $\mathcal T_2$,
and $\mathcal T_3$. Corrections to the
eigenmode first contribute at order $s^4$
in $\lambda_h(s)$, corresponding to order
$s^3$ in $\zeta_{\ed,h}(s)$.

\paragraph*{Dielectric modes.}

The dielectric branches originate from the
kernel of the electrostatic operator.
Let $\mathcal H_\perp=\ker\mathcal T_0$
denote the magnetoquasistatic current space,
whose elements satisfy
\begin{equation}
    \nabla\cdot\mathbf J=0
    \quad\text{in }\Omega,
    \qquad
    \mathbf J\cdot\hat{\mathbf n}=0
    \quad\text{on }\partial\Omega.
    \label{seq:MQS_space}
\end{equation}
Denote by $\mathcal P_\perp$ the orthogonal
projection onto $\mathcal H_\perp$.
Since $\mathcal T_0$ vanishes on this
subspace, the leading-order eigenvalue
problem is determined by the restriction
of $\mathcal T_2$ to $\mathcal H_\perp$:
\begin{equation}
    \mathcal P_\perp
    \mathcal T_2  \mathcal P_\perp \mathbf j_h^\perp
    =
    \lambda_h^{(2)}\mathbf j_h^\perp,
    \qquad
    \mathbf j_h^\perp\in\mathcal H_\perp.
    \label{seq:MQS_eig}
\end{equation}
Here, $\{\mathbf j_h^\perp\}$ denote the
orthonormal magnetoquasistatic current modes.
For these modes, the surface contribution
to $\mathcal T_2$ vanishes, leaving the
magnetostatic integral operator.

Furthermore, the divergence-free and
tangential boundary conditions imply
\begin{equation}
    \int_\Omega\mathbf j_h^\perp\dV
    =\mathbf 0,
\end{equation}
and hence
$\mathcal T_3\mathbf j_h^\perp=\mathbf 0$.
Consequently, the dielectric eigenvalue
branches have no third-order term.

Continuing the perturbative expansion
to fifth order gives
\begin{subequations}
\begin{align}
    \lambda_h(s)
    &=
    s^2\lambda_h^{(2)}
    +s^4\lambda_h^{(4)}
    +s^5\lambda_h^{(5)}
    +O(s^6),
    \\
    \mathbf j_h(s)
    &=
    \mathbf j_h^\perp+O(s^2),
\end{align}
\label{seq:MQS_pert}
\end{subequations}
where
\begin{equation}
    \lambda_h^{(4)} = \langle \vec{j}_h^\perp, \mathcal{T}_4 \vec{j}_h^\perp\rangle + \langle \vec{j}_h^\perp, \mathcal{T}_2 \vec{j}_h^{(2)}\rangle, \qquad \lambda_h^{(5)} = \langle \vec{j}_h^\perp, \mathcal{T}_5 \vec{j}_h^\perp\rangle.
\end{equation}
Since $\zeta_{\Omega,h}(s)=\lambda_h(s)/s$,
comparison with the low-frequency expansion
of the dielectric radiation admittance,
Eq.~\eqref{seq:Lau0_md}, gives
\begin{equation}
\begin{aligned}
    L_{\Omega,h,0}
    &=\lambda_h^{(2)}, \qquad 
    C_{\Omega,h,0}
    &=-\frac{\lambda_h^{(4)}}
             {(\lambda_h^{(2)})^2},\qquad
    g_{\Omega,h,0}
    &=\frac{\lambda_h^{(5)}}
            {(\lambda_h^{(2)})^2}.
\end{aligned}
\label{seq:MQS_parameters}
\end{equation}
The leading inductive coefficient is therefore
determined entirely by the magnetoquasistatic
mode and $\mathcal T_2$. The fourth-order
coefficient also accounts for perturbative
coupling to electroquasistatic currents,
whereas the fifth-order coefficient determines
the leading magnetic-dipole radiation loss.

}


{
\subsection{Bounds on the Static Parameter of the Radiation Impedance}
\subsubsection{Plasmonic Modes}
We now derive a \emph{sharp} upper bound on the product
$r_{\Omega,h,0}C_{\Omega,h,0}$ for an arbitrary plasmonic
mode supported by $\Omega\subseteq B_a$, where $B_a$ is
a ball of radius $a$ centered at the origin.

Combining Eqs.~\eqref{seq:EQS_lambda23} and
\eqref{seq:EQS_parameters}, we obtain
\begin{equation}
    r_{\Omega,h,0}C_{\Omega,h,0}
    =
    -\frac{\lambda_h^{(3)}}{\lambda_h^{(0)}}
    =
    -\frac{
    \langle\vec j_h^\parallel,
    \mathcal T_3\vec j_h^\parallel\rangle
    }{
    \langle\vec j_h^\parallel,
    \mathcal T_0\vec j_h^\parallel\rangle
    }.
    \label{seq:rC_operator}
\end{equation}
We choose the electroquasistatic mode
$\vec j_h^\parallel$ real and introduce the associated
surface density and dipole moment,
\begin{equation}
    \varrho_{s,h}
    :=
    \vec j_h^\parallel\cdot\vers n,
    \qquad
    \boldsymbol\mu_h
    :=
    \int_\Omega\vec j_h^\parallel\,\dV.
    \label{seq:rho_mu_def}
\end{equation}
Using the expression of $\mathcal T_3$ \eqref{seq:T3def}, we obtain
\begin{equation}
    \langle\vec j_h^\parallel,
    \mathcal T_3\vec j_h^\parallel\rangle
    =
    -\frac{|\boldsymbol\mu_h|^2}
    {6\pi\varepsilon_0a^4c_0^3}.
    \label{seq:l3}
\end{equation}
On the other hand, the electrostatic operator $\mathcal{T}_0$ \eqref{seq:T0def}
 gives
\begin{equation}
    \langle\vec j_h^\parallel,
    \mathcal T_0\vec j_h^\parallel\rangle
    =
    \frac{1}{4\pi\varepsilon_0a}
    \langle\vec j_h^\parallel,\nabla f_h\rangle,
\end{equation}
where
\begin{equation}
    f_h(\vec r)
    :=
    \oint_{\partial\Omega}
    \frac{\varrho_{s,h}(\vec r')}
    {|\vec r-\vec r'|}\,\dS'.
\end{equation}
Using the vector identity
$\nabla\cdot(f_h\vec j_h^\parallel)
=f_h\nabla\cdot\vec j_h^\parallel+
\vec j_h^\parallel\cdot\nabla f_h$,
together with $\nabla\cdot\vec j_h^\parallel=0$
in $\Omega$ and the divergence theorem, we obtain
\begin{equation}
    \langle\vec j_h^\parallel,
    \mathcal T_0\vec j_h^\parallel\rangle
    =
    \frac{\mathcal U_h}{4\pi\varepsilon_0a^4},
    \label{seq:l0}
\end{equation}
where
\begin{equation}
    \mathcal U_h^\|
    :=
    \oint_{\partial\Omega}\oint_{\partial\Omega}
    \frac{
    \varrho_{s,h}(\vec r)
    \varrho_{s,h}(\vec r')
    }{
    |\vec r-\vec r'|
    }
    \,\dS'\dS
    \label{seq:U_electrostatic}
\end{equation}
is the Coulomb quadratic form associated with
the surface density $\varrho_{s,h}$.

Substituting Eqs.~\eqref{seq:l3} and \eqref{seq:l0}
into Eq.~\eqref{seq:rC_operator}, we obtain
\begin{equation}
    r_{\Omega,h,0}C_{\Omega,h,0}
    =
    \frac{2}{3c_0^3}
    \frac{|\boldsymbol\mu_h|^2}{\mathcal U_h}.
    \label{seq:rC_electrostatic}
\end{equation}

To bound this quantity, introduce the electrostatic
potential generated by $\varrho_{s,h}$,
\begin{equation}
    \Phi_h(\vec r)
    :=
    \frac{1}{4\pi}
    \oint_{\partial\Omega}
    \frac{\varrho_{s,h}(\vec r')}
    {|\vec r-\vec r'|}\,\dS'.
    \label{seq:potential_Phi}
\end{equation}
Since
$-\nabla^2\Phi_h=
\varrho_{s,h}\delta_{\partial\Omega}$
in the distributional sense, integration by parts
over $\mathbb R^3$ gives
\begin{equation}
    \mathcal U_h^\|
    =
    4\pi\int_{\mathbb R^3}
    |\nabla\Phi_h|^2\,\dV.
    \label{seq:energy_Phi}
\end{equation}
For an arbitrary constant vector
$\vec e\in\mathbb R^3$, consider the auxiliary potential
\begin{equation}
    \Psi_{\vec e}(\vec r)
    :=
    \begin{cases}
        \vec e\cdot\vec r,
        &|\vec r|\leq a,
        \\[2mm]
        \displaystyle
        a^3\frac{\vec e\cdot\vec r}{|\vec r|^3},
        &|\vec r|>a.
    \end{cases}
    \label{seq:auxiliary_potential}
\end{equation}
This potential is continuous across $\partial B_a$,
harmonic on either side, and decays at infinity.
Direct integration yields
\begin{equation}
    \int_{\mathbb R^3}
    |\nabla\Psi_{\vec e}|^2\,\dV
    =
    4\pi a^3|\vec e|^2.
    \label{seq:energy_Psi}
\end{equation}
Since $\nabla\cdot\vec j_h^\parallel=0$ in $\Omega$,
the divergence theorem also gives
\begin{equation}
    \boldsymbol\mu_h
    =
    \oint_{\partial\Omega}
    \vec r\,\varrho_{s,h}(\vec r)\,\dS.
    \label{seq:mu_surface}
\end{equation}
Moreover, $\Omega\subseteq B_a$ implies
$\Psi_{\vec e}=\vec e\cdot\vec r$ on $\partial\Omega$.
Therefore, using Eq.~\eqref{seq:mu_surface}
and the weak form of the equation for $\Phi_h$,
we obtain
\begin{equation}
    \vec e\cdot\boldsymbol\mu_h
    =
    \oint_{\partial\Omega}
    \varrho_{s,h}\Psi_{\vec e}\,\dS
    =
    \int_{\mathbb R^3}
    \nabla\Phi_h\cdot\nabla\Psi_{\vec e}\,\dV.
    \label{seq:dipole_energy_identity}
\end{equation}
Applying the Cauchy--Schwarz inequality,
together with Eqs.~\eqref{seq:energy_Phi}
and \eqref{seq:energy_Psi}, gives
\begin{equation}
    |\vec e\cdot\boldsymbol\mu_h|^2
    \leq
    \left(
    \int_{\mathbb R^3}
    |\nabla\Phi_h|^2\,\dV
    \right)
    \left(
    \int_{\mathbb R^3}
    |\nabla\Psi_{\vec e}|^2\,\dV
    \right)
   =
    a^3\mathcal U_h|\vec e|^2.
    \label{seq:ineq1}
\end{equation}

Since
$|\vec e\cdot\boldsymbol\mu_h|^2
=|\vec e|^2|\boldsymbol\mu_h|^2\cos^2\theta$,
where $\theta$ is the angle between $\vec e$ and
$\boldsymbol\mu_h$, Eq. \eqref{seq:ineq1} becomes $  |\boldsymbol\mu_h|^2\cos^2\theta
    \leq a^3\mathcal U_h$. Thus, choosing $\theta=0$, we obtain
\begin{equation}
    |\boldsymbol\mu_h|^2
    \leq a^3\mathcal U_h.
    \label{seq:dipole_energy_bound}
\end{equation}

Substituting into Eq.~\eqref{seq:rC_electrostatic}
establishes the bound
\begin{equation}
    \boxed{
    r_{\Omega,h,0}C_{\Omega,h,0}
    \leq
    \frac{2}{3}
    \left(\frac{a}{c_0}\right)^3,
    \qquad
    \Omega\subseteq B_a.
    }
    \label{seq:rC_geometry_bound}
\end{equation}
This bound holds for every plasmonic mode admitting
the regular low-frequency expansion considered above.

Equality in \eqref{seq:rC_geometry_bound} is attained by the electric dipole of a sphere. Indeed, for $\Omega=B_a$, the
electroquasistatic electric-dipole mode is uniform,
$\vec j_{\rm ed}^\parallel=\vec j_0$, and its surface
density $\varrho_{s,\rm ed}=\vec j_0\cdot\vers n$
generates
$\Phi_h=\Psi_{\vec j_0}/3$.
Consequently, the gradients in
Eq.~\eqref{seq:dipole_energy_identity}
are proportional for $\vec e=\vec j_0$,
so equality holds in the Cauchy--Schwarz inequality
and hence in Eq.~\eqref{seq:rC_geometry_bound}.

\subsubsection{Dielectric Modes}
We now derive an analogous geometric bound on the product
$g_{\Omega,h,0}L_{\Omega,h,0}$ for an arbitrary dielectric
mode supported by $\Omega\subseteq B_a$.

Combining the perturbative expressions derived above,
we obtain
\begin{equation}
    g_{\Omega,h,0}L_{\Omega,h,0}
    =
    \frac{\lambda_h^{(5)}}{\lambda_h^{(2)}}
    =
    \frac{
    \langle\vec j_h^\perp,
    \mathcal T_5\vec j_h^\perp\rangle
    }{
    \langle\vec j_h^\perp,
    \mathcal T_2\vec j_h^\perp\rangle
    }.
    \label{seq:gL_operator}
\end{equation}
We choose the magnetoquasistatic mode
$\vec j_h^\perp$ real and introduce its magnetic
dipole moment,
\begin{equation}
    \mathbf m_h
    :=
    \frac{1}{2}
    \int_\Omega
    \vec r\times\vec j_h^\perp(\vec r)\,\dV.
    \label{seq:magnetic_moment}
\end{equation}
Since $\nabla\cdot\vec j_h^\perp=0$ in $\Omega$
and $\vec j_h^\perp\cdot\vers n=0$ on
$\partial\Omega$, the divergence theorem gives
\begin{equation}
    \int_\Omega\vec j_h^\perp\,\dV=0,
    \qquad
    \int_\Omega
    (\vec{r} \otimes\vec{j}_h^\perp+\vec{j}_{h}^\perp\otimes\vec{r})\,\dV=0,
    \label{seq:MQS_identities}
\end{equation}
where $\otimes$ denotes the dyadic product.
Using these identities, we obtain
\begin{equation}
    \int_\Omega\int_\Omega
    |\vec r-\vec r'|^2
    \vec j_h^\perp(\vec r)
    \cdot\vec j_h^\perp(\vec r')
    \,\dV'\dV
    =
    -4|\mathbf m_h|^2.
    \label{seq:MQS_dipole_identity}
\end{equation}
The operator $\mathcal{T}_5$ acts on the magnetoquasistatic current $\vec{j}_h^\perp$ as
\begin{equation}
    \mathcal T_5\vec j_h^\perp(\vec r)
    =
    -\frac{\mu_0}{24\pi ac_0^3}
    \int_\Omega
    |\vec r-\vec r'|^2
    \vec j_h^\perp(\vec r')\,\dV' \qquad \vec{j}_h^\perp \in \ker \mathcal{T}_0.
\end{equation}
Using Eq.~\eqref{seq:MQS_dipole_identity}, we obtain
\begin{equation}
    \lambda_h^{(5)}
    =
    \langle\vec j_h^\perp,
    \mathcal T_5\vec j_h^\perp\rangle
    =
    -\frac{\mu_0}{24\pi a^4c_0^3}
    \int_\Omega\int_\Omega
    |\vec r-\vec r'|^2
    \vec j_h^\perp(\vec r)\cdot
    \vec j_h^\perp(\vec r')
    \,\dV'\dV
    =
    \frac{\mu_0|\mathbf m_h|^2}
    {6\pi a^4c_0^3}.
\label{seq:l5}
\end{equation}
On the other hand, since the normal component
of $\vec j_h^\perp$ vanishes on $\partial\Omega$,
the surface contribution to $\mathcal T_2$
vanishes, and we obtain
\begin{equation}
    \langle\vec j_h^\perp,
    \mathcal T_2\vec j_h^\perp\rangle
    =
    \frac{\mu_0\mathcal U_h^\perp}
    {4\pi a^4},
    \label{seq:l2}
\end{equation}
where
\begin{equation}
    \mathcal U_h^\perp
    :=
    \int_\Omega\int_\Omega
    \frac{
    \vec j_h^\perp(\vec r)\cdot
    \vec j_h^\perp(\vec r')
    }{
    |\vec r-\vec r'|
    }
    \,\dV'\dV
    \label{seq:U_magnetostatic}
\end{equation}
is the magnetostatic quadratic form associated
with $\vec j_h^\perp$.

Substituting Eqs.~\eqref{seq:l5} and
\eqref{seq:l2} into Eq.~\eqref{seq:gL_operator},
we obtain
\begin{equation}
    g_{\Omega,h,0}L_{\Omega,h,0}
    =
    \frac{2}{3c_0^3}
    \frac{|\mathbf m_h|^2}{\mathcal U_h^\perp}.
    \label{seq:gL_magnetostatic}
\end{equation}

To bound this quantity, introduce the
magnetostatic vector potential
\begin{equation}
    \vec A_h(\vec r)
    :=
    \frac{1}{4\pi}
    \int_\Omega
    \frac{\vec j_h^\perp(\vec r')}
    {|\vec r-\vec r'|}\,\dV'.
    \label{seq:potential_A}
\end{equation}
Since $\vec j_h^\perp$ is divergence-free and
tangent to $\partial\Omega$, its extension by zero
is solenoidal in $\mathbb R^3$. Consequently,
\begin{equation}
    \nabla\cdot\vec A_h=0,
    \qquad
    \nabla\times\nabla\times\vec A_h
    =\vec j_h^\perp,
\end{equation}
where the current is understood to vanish outside
$\Omega$. Integration by parts over $\mathbb R^3$
then gives
\begin{equation}
    \mathcal U_h^\perp
    =
    4\pi\int_{\mathbb R^3}
    |\nabla\times\vec A_h|^2\,\dV.
    \label{seq:energy_A}
\end{equation}
For an arbitrary constant vector
$\vec e\in\mathbb R^3$, consider the auxiliary
vector potential
\begin{equation}
    \vec A_{\vec e}(\vec r)
    :=
    \begin{cases}
        \displaystyle
        \frac{1}{2}\vec e\times\vec r,
        &|\vec r|\leq a,
        \\[2mm]
        \displaystyle
        \frac{a^3}{2}
        \frac{\vec e\times\vec r}{|\vec r|^3},
        &|\vec r|>a.
    \end{cases}
    \label{seq:auxiliary_vector_potential}
\end{equation}
This vector potential is continuous across
$\partial B_a$ and decays at infinity.
Its curl is
\begin{equation}
    \nabla\times\vec A_{\vec e}
    =
    \begin{cases}
        \vec e,
        &|\vec r|\leq a,
        \\[2mm]
        \displaystyle
        \frac{a^3}{2|\vec r|^3}
        \left[
        3\hat{\vec r}(\vec e\cdot\hat{\vec r})
        -\vec e
        \right],
        &|\vec r|>a.
    \end{cases}
\end{equation}
Direct integration yields
\begin{equation}
    \int_{\mathbb R^3}
    |\nabla\times\vec A_{\vec e}|^2\,\dV
    =
    2\pi a^3|\vec e|^2.
    \label{seq:energy_Ae}
\end{equation}
Since $\Omega\subseteq B_a$, we have
$\vec A_{\vec e}=\vec e\times\vec r/2$
throughout $\Omega$. Therefore,
\begin{equation}
    \vec e\cdot\mathbf m_h
    =
    \int_\Omega
    \vec j_h^\perp\cdot\vec A_{\vec e}\,\dV
    =
    \int_{\mathbb R^3}
    (\nabla\times\vec A_h)\cdot
    (\nabla\times\vec A_{\vec e})\,\dV.
    \label{seq:magnetic_energy_identity}
\end{equation}
Applying the Cauchy--Schwarz inequality,
together with Eqs.~\eqref{seq:energy_A}
and \eqref{seq:energy_Ae}, gives
\begin{equation}
\begin{aligned}
    |\vec e\cdot\mathbf m_h|^2
    &\leq
    \left(
    \int_{\mathbb R^3}
    |\nabla\times\vec A_h|^2\,\dV
    \right)
    \left(
    \int_{\mathbb R^3}
    |\nabla\times\vec A_{\vec e}|^2\,\dV
    \right)
    =
    \frac{a^3}{2}
    \mathcal U_h^\perp|\vec e|^2.
\end{aligned}
\label{seq:magnetic_ineq1}
\end{equation}
Since
$|\vec e\cdot\mathbf m_h|^2
=|\vec e|^2|\mathbf m_h|^2\cos^2\theta$,
where $\theta$ is the angle between $\vec e$
and $\mathbf m_h$, Eq.~\eqref{seq:magnetic_ineq1}
becomes
\begin{equation}
    |\mathbf m_h|^2\cos^2\theta
    \leq
    \frac{a^3}{2}\mathcal U_h^\perp.
\end{equation}
Thus, choosing $\theta=0$, we obtain
\begin{equation}
    |\mathbf m_h|^2
    \leq
    \frac{a^3}{2}\mathcal U_h^\perp.
    \label{seq:magnetic_energy_bound}
\end{equation}
Substituting into Eq.~\eqref{seq:gL_magnetostatic}
establishes the bound
\begin{equation}
    \boxed{
    g_{\Omega,h,0}L_{\Omega,h,0}
    \leq
    \frac{1}{3}
    \left(\frac{a}{c_0}\right)^3,
    \qquad
    \Omega\subseteq B_a.
    }
    \label{seq:gL_geometry_bound}
\end{equation}
This bound holds for every dielectric mode admitting
the regular low-frequency expansion considered above.

}

\section{Fano integral relations}

We derive the Fano integral relations used to establish
the bounds in the main text. Consider two rational
positive-real \emph{immittances}\footnote{The term
\emph{immittance} refers to either an impedance or an
admittance.} $U(s)$ and $V(s)$. Both functions are analytic in
$\mathbb C_+$ and satisfy real symmetry.

We assume that $V(s)$ admits the low-frequency
expansion
\begin{equation}
    V(s)
    =
    \frac{v_{-1}}{s}
    +v_1s+v_2s^2+O(s^3),
    \qquad s\to0,
    \label{seq:zeta_gen_s0}
\end{equation}
where the coefficients are real, with
$v_{-1}>0$ and $v_2<0$.

Consider the reflection coefficient
\begin{equation}
    \Gamma(s)
    =
    \frac{
        U(s)-V(-s)
    }{
        U(s)+V(s)
    }.
    \label{seq:Gamma_scattering}
\end{equation}
Since $V$ is nonconstant and positive real,
$\Re V(s)>0$ in $\mathbb C_+$.
Consequently, $U(s)+V(s)$ does not vanish there.
Thus, $\Gamma$ is rational and meromorphic in
$\mathbb C_+$: its poles are inherited from
$V(-s)$, while its zeros satisfy
\begin{equation}
    U(s)-V(-s)=0.
    \label{seq:Gamma_zeros}
\end{equation}
On the imaginary axis, real symmetry gives
$V(-j\omega)=V(j\omega)^*$, and therefore
\begin{equation}
    1-|\Gamma(j\omega)|^2
    =
    \frac{
        4\Re U(j\omega)\Re V(j\omega)
    }{
        |U(j\omega)+V(j\omega)|^2
    }
    \geq0.
    \label{seq:Gamma_passivity}
\end{equation}
Hence, $|\Gamma(j\omega)|\leq1$.
Boundary values at removable singularities are
understood by continuity.

We first determine the behavior of $\Gamma$
at the origin. The even part of $V$,
\begin{equation}
  \operatorname{Ev} V(s)
    :=
    \frac{V(s)+V(-s)}{2},
\end{equation}
has the expansion
\begin{equation}
  \operatorname{Ev} V(s)
    =
    v_2s^2+O(s^4),
    \qquad s\to0.
    \label{seq:rho_LF}
\end{equation}
Introducing
\begin{equation}
    W(s)
    :=
    \frac{1}{U(s)+V(s)},
    \label{eq:DrivingPointY}
\end{equation}
Eq.~\eqref{seq:Gamma_scattering} becomes
\begin{equation}
    \Gamma(s)
    =
    1-2 W(s)\, \operatorname{Ev}V(s).
    \label{seq:Gamma_rho}
\end{equation}
Since $U(s)$ is rational and positive real,
it is either regular at the origin or has a
simple pole there with nonnegative residue.
Define
\begin{equation}
    u_{-1}
    :=
    \lim_{s\to0}sU(s)
    \geq0.
    \label{seq:U_residue}
\end{equation}
The pole of $V(s)$ at the origin cannot be
canceled by adding $U(s)$, since $v_{-1}>0$
and $u_{-1}\geq0$. Consequently,
\begin{equation}
    W(s)
    =
    \frac{s}{u_{-1}+v_{-1}}
    +O(s^2),
    \qquad s\to0.
    \label{seq:Y_LF}
\end{equation}
Combining Eqs.~\eqref{seq:rho_LF} and
\eqref{seq:Y_LF} yields
\begin{equation}
    \Gamma(s)
    =
    1-
    \frac{2v_2}{u_{-1}+v_{-1}}s^3
    +O(s^4),
    \qquad s\to0.
    \label{seq:Gamma_LF}
\end{equation}
The origin is therefore a removable singularity
of $\Gamma$, with $\Gamma(0)=1$.
Choosing the local logarithm to vanish at the
origin gives
\begin{equation}
    \log\frac{1}{\Gamma(s)}
    =
    \frac{2v_2}{u_{-1}+v_{-1}}s^3
    +O(s^4).
    \label{seq:logGamma_LF}
\end{equation}

To construct an analytic logarithm throughout
$\mathbb C_+$, we remove the right-half-plane
zeros and poles of $\Gamma$.
Following the factorization used in broadband
matching theory~\cite{youla_new_1964},
let $\{z_k\}$ and $\{p_\ell\}$ denote its
right-half-plane zeros and poles, respectively,
counted with multiplicity.

Define the normalized all-pass factor
\begin{equation}
    A(q,s)
    :=
    -\frac{q^*}{q}
    \frac{s-q}{s+q^*},
    \qquad q\in\mathbb C_+,
\end{equation}
and the finite products
\begin{equation}
    B(s)
    :=
    \prod_{z_k\in\mathbb C_+}A(z_k,s),
    \qquad
    \Sigma(s)
    :=
    \prod_{p_\ell\in\mathbb C_+}A(p_\ell,s).
    \label{seq:allpass_factors}
\end{equation}
An empty product is understood to be unity.
These factors satisfy
\begin{equation}
    B(0)=\Sigma(0)=1,
    \qquad
    |B(j\omega)|=|\Sigma(j\omega)|=1.
\end{equation}
Dividing by $B$ removes the right-half-plane
zeros, while multiplying by $\Sigma$ cancels
the right-half-plane poles.
The resulting function
\begin{equation}
    \Gamma_{\rm mp}(s)
    :=
    B(s)^{-1}\Sigma(s)\Gamma(s)
\end{equation}
is analytic and zero-free in $\mathbb C_+$, and thus is \emph{minimum-phase}.
Moreover, $\Gamma_{\rm mp}(0)=1$ and
\begin{equation}
    |\Gamma_{\rm mp}(j\omega)|
    =
    |\Gamma(j\omega)|.
    \label{seq:Gamma_boundary}
\end{equation}

Since $\mathbb C_+$ is simply connected,
$\Gamma_{\rm mp}$ admits an analytic logarithm.
Define
\begin{equation}
    F(s)
    :=
    \log\frac{1}{\Gamma_{\rm mp}(s)},
    \label{seq:F_definition}
\end{equation}
choosing the branch whose analytic continuation
to the origin satisfies $F(0)=0$.
This normalization preserves real symmetry:
$F(s^*)=F(s)^*$.

Near the origin, the local logarithms normalized
to vanish at $s=0$ satisfy
\begin{equation}
    F(s)
    =
    \log\frac{1}{\Gamma(s)}
    -\log\Sigma(s)
    +\log B(s).
    \label{seq:F_local}
\end{equation}
This decomposition is used locally; the analytic
logarithm throughout $\mathbb C_+$ is defined
by Eq.~\eqref{seq:F_definition}.

The zeros and poles of $\Gamma$ occur in
conjugate pairs, so $B$ and $\Sigma$ satisfy
real symmetry. Moreover,
$A(q,-s)=A(q^*,s)^{-1}$ implies
\begin{equation}
    B(s)B(-s)=1,
    \qquad
    \Sigma(s)\Sigma(-s)=1.
\end{equation}
Both factors are analytic and nonzero near
the origin. Their local logarithms, normalized
by $\log B(0)=\log\Sigma(0)=0$, are therefore
odd and have real Taylor coefficients:
\begin{subequations}
\label{seq:B_Sigma_expansions}
\begin{align}
    \log B(s)
    &=
    b_1s+b_3s^3+O(s^5),
    \\
    \log\Sigma(s)
    &=
    \sigma_1s+\sigma_3s^3+O(s^5).
\end{align}
\end{subequations}
Explicitly,
\begin{equation}
    b_m
    =
    -\frac{2}{m}
    \sum_{z_k \in \mathbb{C}_+} \Re z_k^{-m} ,
    \qquad
    \sigma_m
    =
    -\frac{2}{m}
    \sum_{p_\ell \in \mathbb{C}_+}\Re p_\ell^{-m},
    \qquad m=1,3.
    \label{seq:allpass_coefficients}
\end{equation}
Using Eqs.~\eqref{seq:logGamma_LF} and
\eqref{seq:F_local}, we obtain
\begin{equation}
    F(s)
    =
    a_1s+a_3s^3+O(s^4),
    \qquad s\to0,
    \label{seq:F_LF}
\end{equation}
where
\begin{subequations}
\begin{align}
    a_1
    &=
    b_1-\sigma_1,
    \label{seq:a1}
    \\
    a_3
    &=
    \frac{2v_2}{u_{-1}+v_{-1}}
    +b_3-\sigma_3.
    \label{seq:a3}
\end{align}
\end{subequations}
Furthermore, Eqs.~\eqref{seq:Gamma_boundary}
and \eqref{seq:F_definition} give
\begin{equation}
    \Re F(j\omega)
    =
    \ln\frac{1}{|\Gamma(j\omega)|}.
    \label{seq:ReFjw}
\end{equation}
Since $a_1$ and $a_3$ are real,
Eq.~\eqref{seq:F_LF} implies
$\Re F(j\omega)=O(\omega^4)$ as
$\omega\to0$.
Thus, the weighted integrals derived below
converge at the origin.

\subsubsection{Weighted Fano Integrals}

\begin{figure}
    \centering
    \includegraphics[width=0.25\linewidth]{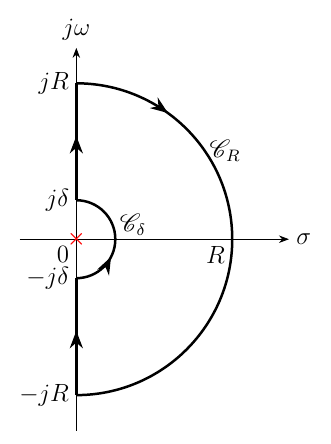}
    \caption{
    Right-half-plane contour $\mathscr C$ used in the derivation of the
    Fano integral relations. The origin is excluded by the semicircular
    indentation $\mathscr C_\delta$.
    }
    \label{fig:contour}
\end{figure}

We now derive the weighted Fano integral relations.
Consider the half-annular domain
\begin{equation}
    D
    =
    \left\{
    s\in\mathbb C_+:
    \delta<|s|<R
    \right\},
    \qquad 0<\delta<R,
\end{equation}
and its clockwise-oriented boundary
$\mathscr C$, shown in Fig.~\ref{fig:contour}.
The imaginary-axis segments are traversed upward,
while the inner semicircle $\mathscr C_\delta$
runs from $-j\delta$ to $j\delta$ through
$\mathbb C_+$.

Since $F(s)/s^2$ is analytic in $D$,
Cauchy's theorem gives
\begin{equation}
    \oint_{\mathscr C}
    \frac{F(s)}{s^2}\,\mathrm ds
    =
    0.
    \label{seq:contour_I2}
\end{equation}
Using real symmetry to combine the
imaginary-axis segments, we obtain
\begin{equation}
\begin{aligned}
    -2j\int_\delta^R
    \frac{\Re F(j\omega)}{\omega^2}
    \,\mathrm d\omega
    +
    \int_{\mathscr C_\delta}
    \frac{F(s)}{s^2}\,\mathrm ds
    +
    \int_{\mathscr C_R}
    \frac{F(s)}{s^2}\,\mathrm ds
    =0.
\end{aligned}
\end{equation}
The large-semicircle integral vanishes as
$R\to\infty$, since $F(s)=O(\log R)$
there. From Eq.~\eqref{seq:F_LF},
the inner semicircle contributes
\begin{equation}
    \lim_{\delta\to0}
    \int_{\mathscr C_\delta}
    \frac{F(s)}{s^2}\,\mathrm ds
    =
    j\pi a_1.
\end{equation}
Taking the limits and using
Eqs.~\eqref{seq:a1} and \eqref{seq:ReFjw}
gives the $\omega^{-2}$-weighted
Fano relation,
\begin{equation}
    \boxed{
    \frac{2}{\pi}
    \int_0^\infty
    \ln\frac{1}{|\Gamma(j\omega)|}
    \frac{\mathrm d\omega}{\omega^2}
    =
    b_1-\sigma_1.
    }
    \label{seq:Fano_I2}
\end{equation}
To derive the next relation, subtract the
linear term of $F$:
\begin{equation}
    F_3(s):=F(s)-a_1s.
    \label{seq:F3}
\end{equation}
Since $a_1$ is real,
\begin{equation}
    \Re F_3(j\omega)
    =
    \Re F(j\omega),
\end{equation}
while the low-frequency expansion becomes
\begin{equation}
    F_3(s)
    =
    a_3s^3+O(s^4).
\end{equation}
Applying Cauchy's theorem to
$F_3(s)/s^4$ over the same contour gives
\begin{equation}
\begin{aligned}
    2j\int_\delta^R
    \frac{\Re F(j\omega)}{\omega^4}
    \,\mathrm d\omega
    +
    \int_{\mathscr C_\delta}
    \frac{F_3(s)}{s^4}\,\mathrm ds
    +
    \int_{\mathscr C_R}
    \frac{F_3(s)}{s^4}\,\mathrm ds
    =0.
\end{aligned}
\end{equation}
The large-semicircle contribution vanishes
as $R\to\infty$, while
\begin{equation}
    \lim_{\delta\to0}
    \int_{\mathscr C_\delta}
    \frac{F_3(s)}{s^4}\,\mathrm ds
    =
    j\pi a_3.
\end{equation}
Using Eq.~\eqref{seq:a3}, we obtain the
$\omega^{-4}$-weighted Fano relation,
\begin{equation}
    \boxed{
    \frac{2}{\pi}
    \int_0^\infty
    \ln\frac{1}{|\Gamma(j\omega)|}
    \frac{\mathrm d\omega}{\omega^4}
    =
    -\frac{2v_2}{u_{-1}+v_{-1}}
    +\sigma_3-b_3.
    }
    \label{seq:Fano_I4}
\end{equation}
These relations connect the weighted
logarithmic mismatch to the low-frequency
immittance coefficients and the right-half-plane
pole--zero structure of the reflection coefficient.
In particular, when $\Gamma$ has a single real
right-half-plane pole, they yield a bound
independent of the locations of its zeros,
as established by the following corollary.

\begin{corollary}
\label{cor:bound}
Let $U(s)$ and $V(s)$ be rational positive-real
immittances satisfying the preceding assumptions.
Suppose that, besides its pole at the origin,
$V(s)$ has exactly one additional pole,
which is simple and located at $s=-p$,
with $p>0$.
Then the reflection coefficient defined in
Eq.~\eqref{seq:Gamma_scattering} satisfies
\begin{equation}
    \boxed{
    \int_0^\infty
    \ln\frac{1}{|\Gamma(j\omega)|}
    \frac{\mathrm d\omega}{\omega^4}
    \leq
    -\pi\frac{v_2}{v_{-1}}.
    }
    \label{seq:Fano_single_pole_bound}
\end{equation}
\end{corollary}

\begin{proof}
Since $U(s)+V(s)$ is analytic and nonzero in
$\mathbb C_+$, the reflection coefficient
has exactly one simple pole in $\mathbb C_+$,
located at $s=p$.
Its normalized all-pass pole factor is
\begin{equation}
    \Sigma(s)=\frac{p-s}{p+s},
\end{equation}
whose logarithmic coefficients satisfy
\begin{equation}
    \sigma_1=-\frac{2}{p},
    \qquad
    \sigma_3=-\frac{2}{3p^3}
    =\frac{\sigma_1^3}{12}.
    \label{seq:single_pole_coefficients}
\end{equation}
Let $\{z_k\}$ denote the right-half-plane
zeros of $\Gamma$, counted with multiplicity.
Writing $z_k^{-1}=x_k+jy_k$, with $x_k>0$,
we obtain
\begin{equation}
\begin{aligned}
    b_3
    =
    -\frac{2}{3}
    \sum_k\Re(z_k^{-3})
    =
    -\frac{2}{3}
    \sum_k(x_k^3-3x_ky_k^2)
    \geq
    -\frac{2}{3}
    \left(\sum_kx_k\right)^3
    =
    \frac{b_1^3}{12}.
\end{aligned}
\label{seq:single_pole_zero_bound}
\end{equation}
On the other hand, passivity and
Eq.~\eqref{seq:Fano_I2} imply
$b_1\geq\sigma_1$.
Since the cubic function is monotonically
increasing on the real axis,
\begin{equation}
    b_3
    \geq
    \frac{b_1^3}{12}
    \geq
    \frac{\sigma_1^3}{12}
    =
    \sigma_3.
\end{equation}
Thus, $\sigma_3-b_3\leq0$, and
Eq.~\eqref{seq:Fano_I4} yields
\begin{equation}
    \int_0^\infty
    \ln\frac{1}{|\Gamma(j\omega)|}
    \frac{\mathrm d\omega}{\omega^4}
    \leq
    -\frac{\pi v_2}{u_{-1}+v_{-1}}.
\end{equation}
Finally, since $u_{-1}\geq0$,
$v_{-1}>0$, and $v_2<0$, we have
\begin{equation}
    -\frac{v_2}{u_{-1}+v_{-1}}
    \leq
    -\frac{v_2}{v_{-1}},
\end{equation}
which proves the assertion.
\end{proof}

\subsection{Branch-cut contributions to the Fano integral relations}

We now extend the preceding Fano integral relations ~\eqref{seq:Fano_I2} and \eqref{seq:Fano_I4} to impedances whose analytic continuation exhibits branch points. We retain the positive-realness and low-frequency assumptions used above, but allow $\zeta(s)$ to be nonrational.

\begin{figure}
    \centering
    \includegraphics[width=0.5\linewidth]{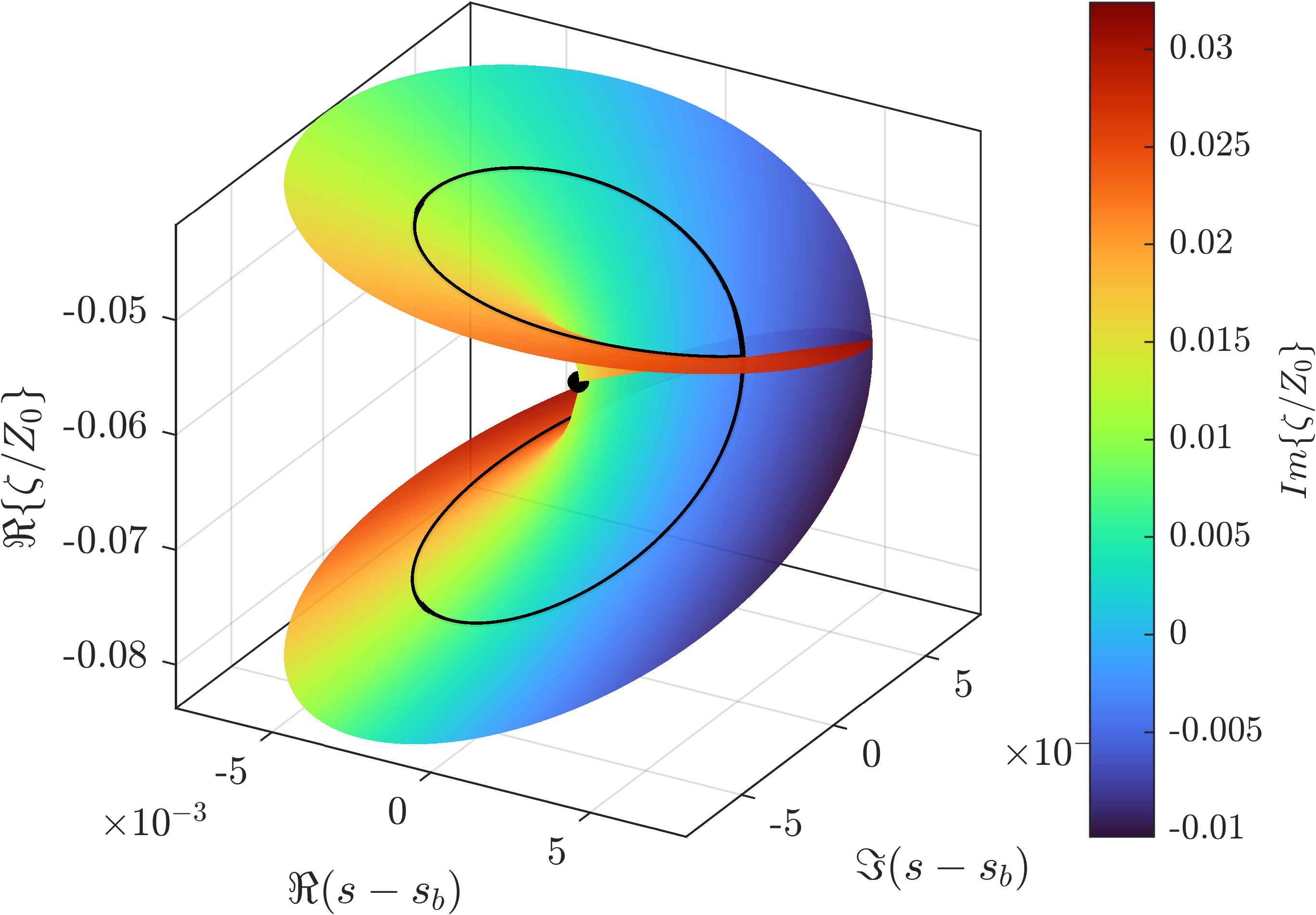}
    \caption{
    Riemann surface of the electric-dipole radiation
    impedance of a sphere near a branch point
    $s_b\in\mathbb C_-$ (black marker).
    The surface height and color represent
    $\Re\{\zeta/Z_0\}$ and $\Im\{\zeta/Z_0\}$, respectively.
    The black curve traces two successive circuits around
    $s_b$: the first reaches the second sheet, while the
    second returns to the original sheet, illustrating
    the square-root character of the branch point.
    }
    \label{fig:riemann}
\end{figure}

Such branch points may occur in the analytic continuation
of the full-wave radiation impedance. In particular,
certain spectral degeneracies of the radiation operator
connect distinct eigenvalue branches. Analytic continuation
around these points can interchange the branches, thereby
connecting different sheets of the associated Riemann
surface, as illustrated in Fig.~\ref{fig:riemann}.

\begin{figure}
    \centering
    \includegraphics[width=0.85\linewidth]{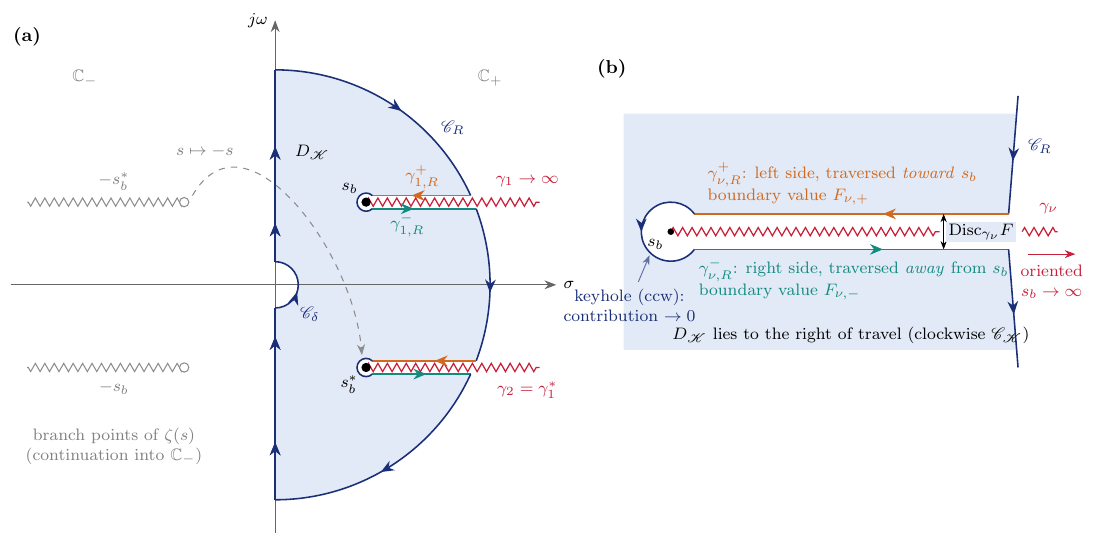}
\caption{Integration contour for the Fano relations in the presence of branch cuts.
(a)~Branch points of $\zeta(s)$ in $\mathbb{C}_-$ (grey) are mapped by $s\mapsto -s$
to branch points $s_b, s_b^{*}\in\mathbb{C}_+$ of the reflected impedance $\zeta(-s)$,
and hence of $\Gamma$. The conjugate-symmetric cuts $\gamma_\nu$ (red) run from the branch
points to infinity, avoiding the imaginary axis and the origin. The clockwise boundary
$\mathscr{C}_{\mathscr K}$ of $D_{\mathscr K}=D\setminus\mathscr K$ consists of the imaginary
axis, indented at the origin ($\mathscr C_\delta$);
the outer arc $\mathscr C_R$; and the two sides of each truncated cut. 
(b)~Detail of one cut. The left side $\gamma_{\nu,R}^{+}$ is traversed toward $s_b$, and the
right side $\gamma_{\nu,R}^{-}$ away from it. Their contributions combine into the integral of
the jump $\operatorname{Disc}_{\gamma_\nu}F=F_{\nu,+}-F_{\nu,-}$. The keyhole contribution vanishes because $F$ has at most a logarithmic singularity at $s_b$.}    \label{fig:IntegralBranchCut}
\end{figure}

Although $\zeta(s)$ is analytic in $\mathbb C_+$ on the
physical sheet \footnote{The physical sheet is the sheet of the Riemann surface containing the impedance branch associated with the chosen mode of the causal scattering problem with outgoing radiation conditions. This branch is identified
by continuation from its quasistatic limit. Its physical frequency response is obtained from the right half-plane
as $\zeta(j\omega)=\lim_{\sigma\to0^+}\zeta(\sigma+j\omega)$.}, its reflected continuation $\zeta(-s)$
may have branch points in $\mathbb C_+$. These singularities can be inherited by the reflection coefficient $\Gamma(s)$ defined in Eq.~\eqref{seq:Gamma_scattering}, as sketched in Fig. \ref{fig:IntegralBranchCut}.

Choose a fixed system of nonintersecting branch cuts
\begin{equation}
    \mathscr K
    :=
    \bigcup_{\nu=1}^{N_c}\gamma_\nu,
\end{equation}
each extending from a branch point to infinity.

The cut system is invariant under complex conjugation and avoids the imaginary axis and a neighborhood of the origin. We assume that $\mathbb C_+\setminus\mathscr K$ is simply connected and that the selected continuation of $\Gamma$ is single-valued and meromorphic there.

Within the truncated domain
$D=\{s\in\mathbb C_+:\delta<|s|<R\}$, define
\begin{equation}
    D_{\mathscr K}
    :=
    D\setminus\mathscr K,
    \label{seq:cut_domain}
\end{equation}
and denote the truncated cuts by
$\gamma_{\nu,R}:=\gamma_\nu\cap D$.
The contours are chosen so that $D_{\mathscr K}$ is
simply connected. Using fixed cuts and then truncating
them ensures a consistent choice of sheet as $R$ varies.

Assume that the isolated zeros and poles of $\Gamma$ on
the selected sheet admit normalized all-pass factors
$B$ and $\Sigma$, respectively, with the real-symmetry
and boundary-modulus properties used above.
For infinite zero or pole sets, the products are assumed
to converge locally uniformly and to admit the local
expansions in Eq.~\eqref{seq:B_Sigma_expansions}.
In particular, both factors are analytic and nonzero
near the origin, with $B(0)=\Sigma(0)=1$.

The regularized reflection coefficient
\begin{equation}
    \widetilde{\Gamma}(s)
    :=
    B(s)^{-1}\Sigma(s)\Gamma(s)
\end{equation}
is then analytic and zero-free in
$\mathbb C_+\setminus\mathscr K$, while retaining the
branch discontinuities across the cuts.
Since this domain is simply connected, we can define
\begin{equation}
    F(s)
    :=
    \log\frac{1}{\widetilde{\Gamma}(s)},
\end{equation}
choosing the analytic logarithm whose continuation to
the origin satisfies $F(0)=0$.
The local expansion remains
\begin{equation}
    F(s)
    =
    a_1s+a_3s^3+O(s^4),
\end{equation}
with
\begin{equation}
    a_1=b_1-\sigma_1,
    \qquad
    a_3=2\frac{v_2}{u_{-1}+v_{-1}}+b_3-\sigma_3,
\end{equation}
as in Eqs.~\eqref{seq:F_LF} and \eqref{seq:a1} and \eqref{seq:a3}  .
Real symmetry and the all-pass property also give
\begin{equation}
    F(s^*)=F(s)^*,
    \qquad
    \Re F(j\omega)
    =
    \ln\frac{1}{|\Gamma(j\omega)|}.
\end{equation}

Orient each cut from its branch point toward infinity,
and denote its left and right sides by $+$ and $-$,
respectively. Define the \textit{discontinity} jump
\begin{equation}
    \operatorname{Disc}_{\gamma_\nu}F(s)
    :=
    F_{\nu,+}(s)-F_{\nu,-}(s).
    \label{seq:DiscF}
\end{equation}
Both boundary values are obtained by continuation of
the same logarithm normalized by $F(0)=0$.

For the limiting contour argument, we assume that $F$
has at most logarithmic singularities at the finite
branch points. This condition includes algebraic
branch points of $\widetilde{\Gamma}$ and ensures that
the corresponding indentation contributions vanish.
We also assume absolute convergence of the weighted
jump integrals appearing below and the sufficient
growth condition
\begin{equation}
    F(s)=o(|s|),
    \qquad |s|\to\infty,
    \label{seq:F_branch_growth}
\end{equation}
uniformly along the outer semicircles on the selected
sheet. These assumptions replace the automatic
large-frequency control provided by rationality.
An infinite, locally finite cut system can be treated
in the same way if the sums of the weighted jump
integrals also converge absolutely.

Let $\mathscr C_{\mathscr K}$ denote the
clockwise-oriented boundary of $D_{\mathscr K}$.
Since $F(s)/s^2$ is analytic in this domain,
\begin{equation}
    \oint_{\mathscr C_{\mathscr K}}
    \frac{F(s)}{s^2}\,\mathrm ds
    =
    0.
    \label{seq:contour_branch_I2}
\end{equation}
After shrinking the indentations around the branch
points, the boundary consists of the original contour
$\mathscr C$ and the two sides of each truncated cut:
\begin{equation}
    \mathscr C_{\mathscr K}
    =
    \mathscr C
    +
    \sum_{\nu=1}^{N_c}
    \left(
        \gamma_{\nu,R}^{+}
        +
        \gamma_{\nu,R}^{-}
    \right).
    \label{seq:cut_contour_decomposition}
\end{equation}
Here, integrals over $\mathscr C$ are evaluated
piecewise using the selected sheet.
Any isolated imaginary-axis zeros are treated by
vanishing indentations, as in the preceding derivation.

With clockwise boundary orientation, the left side of
each cut is traversed toward the branch point, while
the right side is traversed toward the outer boundary.
Consequently,
\begin{equation}
    \int_{\gamma_{\nu,R}^{+}}
    \frac{F(s)}{s^2}\,\mathrm ds
    +
    \int_{\gamma_{\nu,R}^{-}}
    \frac{F(s)}{s^2}\,\mathrm ds
    =
    -\int_{\gamma_{\nu,R}}
    \frac{\operatorname{Disc}_{\gamma_\nu}F(s)}
         {s^2}\,\mathrm ds.
    \label{seq:cut_jump_relation}
\end{equation}
Substitution into Eq.~\eqref{seq:contour_branch_I2}
therefore gives
\begin{equation}
    \oint_{\mathscr C}
    \frac{F(s)}{s^2}\,\mathrm ds
    =
    \sum_{\nu=1}^{N_c}
    \int_{\gamma_{\nu,R}}
    \frac{\operatorname{Disc}_{\gamma_\nu}F(s)}
         {s^2}\,\mathrm ds.
    \label{seq:contour_branch_jump}
\end{equation}

The growth condition~\eqref{seq:F_branch_growth}
makes the outer-semicircle contribution vanish.
Taking $\delta\to0$ and $R\to\infty$, and evaluating
the imaginary-axis and origin contributions as before,
we obtain
\begin{equation}
    -2j\int_0^\infty
    \frac{\Re F(j\omega)}{\omega^2}\,\mathrm d\omega
    +j\pi a_1
    =
    \sum_{\nu=1}^{N_c}
    \int_{\gamma_\nu}
    \frac{\operatorname{Disc}_{\gamma_\nu}F(s)}
         {s^2}\,\mathrm ds.
\end{equation}
Hence, the first generalized Fano relation is
\begin{equation}
    \boxed{
    \frac{2}{\pi}
    \int_0^\infty
    \ln\frac{1}{|\Gamma(j\omega)|}
    \frac{\mathrm d\omega}{\omega^2}
    =
    b_1-\sigma_1
    -
    \frac{1}{j\pi}
    \sum_{\nu=1}^{N_c}
    \int_{\gamma_\nu}
    \frac{\operatorname{Disc}_{\gamma_\nu}F(s)}
         {s^2}\,\mathrm ds.
    }
    \label{seq:Fano_I2_branch}
\end{equation}

For the second relation, define
$F_3(s)=F(s)-a_1s$ as before.
Since the subtracted term is single-valued,
\begin{equation}
    \operatorname{Disc}_{\gamma_\nu}F_3(s)
    =
    \operatorname{Disc}_{\gamma_\nu}F(s).
\end{equation}
Moreover, $\Re F_3(j\omega)=\Re F(j\omega)$ and
$F_3(s)=a_3s^3+O(s^4)$ near the origin.
Applying Cauchy's theorem to $F_3(s)/s^4$ yields
\begin{equation}
    \oint_{\mathscr C}
    \frac{F_3(s)}{s^4}\,\mathrm ds
    =
    \sum_{\nu=1}^{N_c}
    \int_{\gamma_{\nu,R}}
    \frac{\operatorname{Disc}_{\gamma_\nu}F(s)}
         {s^4}\,\mathrm ds.
    \label{seq:contour_branch_I4}
\end{equation}
Equation~\eqref{seq:F_branch_growth} implies
$F_3(s)=O(|s|)$ at infinity, so the outer-semicircle
contribution again vanishes. Taking the limits gives
\begin{equation}
    2j\int_0^\infty
    \frac{\Re F(j\omega)}{\omega^4}\,\mathrm d\omega
    +j\pi a_3
    =
    \sum_{\nu=1}^{N_c}
    \int_{\gamma_\nu}
    \frac{\operatorname{Disc}_{\gamma_\nu}F(s)}
         {s^4}\,\mathrm ds.
\end{equation}
Therefore,
\begin{equation}
    \boxed{
    \frac{2}{\pi}
    \int_0^\infty
    \ln\frac{1}{|\Gamma(j\omega)|}
    \frac{\mathrm d\omega}{\omega^4}
    =
     -\frac{2v_2}{u_{-1}+v_{-1}}+\sigma_3-b_3
    +
    \frac{1}{j\pi}
    \sum_{\nu=1}^{N_c}
    \int_{\gamma_\nu}
    \frac{\operatorname{Disc}_{\gamma_\nu}F(s)}
         {s^4}\,\mathrm ds.
    }
    \label{seq:Fano_I4_branch}
\end{equation}

Conjugation exchanges the left and right sides of each
cut. Together with real symmetry, this implies that
the total jump integrals in
Eqs.~\eqref{seq:Fano_I2_branch} and
\eqref{seq:Fano_I4_branch} are purely imaginary, so
the corresponding corrections are real.

Under the stated factorization and convergence assumptions,
these equations extend the preceding Fano relations to
impedances with branch points. The previous relations
are recovered when the cut contributions vanish.
Extending the rational-model bound to the full-wave
problem therefore requires control of these additional
contributions.

\subsection{Small-particle Bode--Fano bound for the electric dipole}

We now specialize the preceding framework to
electric-dipole ($h=\mathrm{ed}$) scattering in the
small-particle limit. Throughout this subsection,
we omit the modal index $h$, with all quantities
understood to refer to the electric-dipole mode.

Under assumption~(H3), the radiation impedance
$\zeta_\Omega(s)$ is approximated by the rational
positive-real function $\zeta_\Omega^{\rm sp}(s)$
defined in Eq.~\eqref{eq:zeta_qs}. We assume that
the material impedance $Z_\chi(s)$ is also rational
and positive real, and introduce
\begin{equation}
    \Gamma_{\rm sp}(s)
    :=
    \frac{
        Z_\chi(s)-\zeta_\Omega^{\rm sp}(-s)
    }{
        Z_\chi(s)+\zeta_\Omega^{\rm sp}(s)
    }.
    \label{seq:Gamma_sm}
\end{equation}
Besides its pole at the origin,
$\zeta_\Omega^{\rm sp}$ has exactly one simple pole
at $s=-p$, where
$p=R_{\Omega,0}/L_{\Omega,0}>0$.
Thus, the hypotheses of the Corollary \ref{cor:bound}
are satisfied. Identifying $U=Z_    \chi$, $V = \zsp$, $-v_2=r_{\Omega,0}$, $v_{-1}=C^{-1}_{\Omega,0}$
Eq.~\eqref{seq:Fano_single_pole_bound} gives
\begin{equation}
    \int_0^\infty
    \ln\frac{1}{|\Gamma_{\rm sp}(j\omega)|}
    \frac{\mathrm d\omega}{\omega^4}
    \leq
    \pi r_{\Omega,0}C_{\Omega,0}.
    \label{seq:BF_ed_intermediate}
\end{equation}
Combining \eqref{seq:BF_ed_intermediate} with  the  bound~\eqref{seq:rC_geometry_bound}, yields the small-particle Bode--Fano bound
\begin{equation}
    \boxed{
    \int_0^\infty
    \ln\frac{1}{|\Gamma_{\rm sp}(j\omega)|}
    \frac{\mathrm d\omega}{\omega^4}
    \leq
    \frac{2\pi}{3}
    \left(\frac{a}{c_0}\right)^3,
    \qquad
    \Omega\subseteq B_a.
    }
    \label{seq:BF_geometric_bound}
\end{equation}

Since the integrand is nonnegative, this bound
also constrains the reflection coefficient over
any finite frequency band
$\mathcal B=[\omega_1,\omega_2]$.
Assume that
$|\Gamma_{\rm sp}(j\omega)|\leq\Gamma_{\max}<1$
throughout $\mathcal B$. Then
\begin{equation}
    \ln\frac{1}{\Gamma_{\max}}
    \int_{\omega_1}^{\omega_2}
    \frac{\mathrm d\omega}{\omega^4}
    \leq
    \pi r_{\Omega,0}C_{\Omega,0}.
\end{equation}
Introducing the center frequency
$\omega_0=\sqrt{\omega_1\omega_2}$ and the bandwidth
$\Delta \omega= \omega_2-\omega_1$,
direct integration gives
\begin{equation}
    \ln\frac{1}{\Gamma_{\max}}
 \frac{\Delta \omega}{\omega_0}
    \left[1+\frac{1}{3}\left(\frac{\Delta \omega}{\omega_0}\right)^2\right]
    \leq
    \frac{\pi}{Q_{\Omega,0}},
    \label{seq:BF_band_exact}
\end{equation}
where
\begin{equation}
    Q_{\Omega,0}
    :=
    \frac{1}{
    \omega_0^3r_{\Omega,0}C_{\Omega,0}
    } 
\end{equation}
is the radiation quality factor.
Since the factor in the square brackets in Eq. \eqref{seq:BF_band_exact} is not smaller
than unity, we finally obtain
\begin{equation}
    \boxed{
    \ln\frac{1}{\Gamma_{\max}}
    \frac{\Delta \omega}{\omega_0}
    \leq
    \frac{\pi}{Q_{\Omega,0}}
    }
    \label{seq:BF_band_Q}
\end{equation}
with $k_0=\omega_0/c_0$.

{

\subsection{Small-particle Bode--Fano bound for the magnetic dipole}

We now specialize the preceding framework to
magnetic-dipole ($h=\mathrm{md}$) scattering in the
small-particle limit. Throughout this subsection,
we omit the modal index $h$, with all quantities
understood to refer to the magnetic-dipole mode.

Let $Y_\chi(s)=1/Z_\chi(s)$ denote the material
admittance, and let $Y_\Omega^{\rm sp}(s)$ be the
small-particle radiation admittance defined in
Eq.~\eqref{eq:zeta_qs_diel}.
We introduce the corresponding reflection coefficient
\begin{equation}
    \Gamma_{\rm sp}(s)
    =
    \frac{
        Y_\chi(s)-Y_\Omega^{\rm sp}(-s)
    }{
        Y_\chi(s)+Y_\Omega^{\rm sp}(s)
    }.
    \label{seq:Gamma_md}
\end{equation}
On the imaginary axis, this coefficient has the same
magnitude as its impedance-based counterpart.

Besides the pole at the origin, the small-particle
radiation admittance has exactly one simple pole at
$s=-p$, where
$p=C_{\Omega,0}/g_{\Omega,0}>0$.
Since both admittances are rational and positive real,
the hypotheses of the single-pole corollary are satisfied.
Identifying $U = Y_\chi$, $V=Y_{\Omega}^{\rm sp}$ $-v_2=g_{\Omega,0}$ and $v_{-1}=L^{-1}_{\Omega,0}$,
Eq.~\eqref{seq:Fano_single_pole_bound} yields
\begin{equation}
    \int_0^\infty
    \ln\frac{1}{|\Gamma_{\rm sp}(j\omega)|}
    \frac{\mathrm d\omega}{\omega^4}
    \leq
    \pi g_{\Omega,0}L_{\Omega,0}.
    \label{seq:BF_md}
\end{equation}

Combining \eqref{seq:BF_md} with the bound
\eqref{seq:gL_geometry_bound} yields the small-particle Bode--Fano bound
\begin{equation}
    \boxed{
    \int_0^\infty
    \ln\frac{1}{|\Gamma_{\rm sp}(j\omega)|}
    \frac{\mathrm d\omega}{\omega^4}
    \leq
    \frac{\pi}{3}
    \left(\frac{a}{c_0}\right)^3,
    \qquad
    \Omega\subseteq B_a.
    }
    \label{seq:BF_md_geometry}
\end{equation}

}

\section{Network Synthesis}

\subsection{Frequency Transformations}

We now establish the connection between passive material
dispersion and the normalized low-pass prototypes used
for network synthesis.

Consider an arbitrary rational positive-real material
impedance $Z_\chi(s)$. By Darlington's synthesis theorem,
it can be realized as the driving-point impedance of a
lossless two-port network terminated by a resistance
$R_m>0$.

In particular, the high-frequency inductive contribution
can be extracted from the lossless network, yielding
\begin{equation}
    Z_\chi(s)
    =
    sL_{\chi,\infty}+Z_\chi'(s),
    \qquad
    L_{\chi,\infty}
    :=
    \lim_{s\to\infty}\frac{Z_\chi(s)}{s}
    \geq0,
    \label{seq:material_inductance_extraction}
\end{equation}
where $Z_\chi'(s)$ remains rational and positive real
and admits a Darlington realization consisting of a
lossless two-port network terminated by an arbitrary resistor $R_m$\footnote{The value $R_m$ can be chosen freely by incorporating an ideal transformer in the lossless network.}. 
Thus, material dissipation is represented by the
terminating resistance, whereas its dispersive
properties are synthesized by the lossless network.

\begin{figure}
    \centering
    \includegraphics[width=1\linewidth]{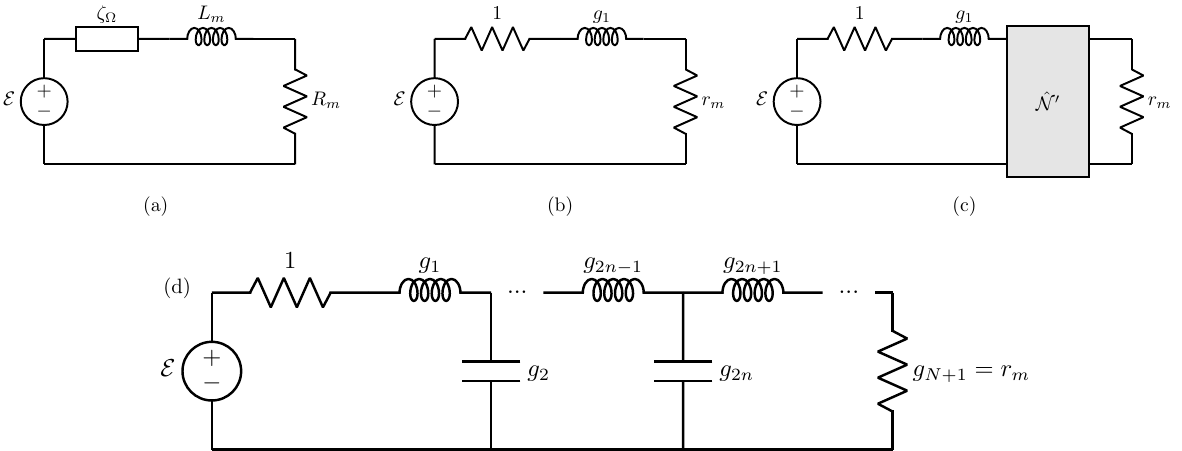}
  \caption{
    Network representation of passive material-dispersion synthesis.
    (a) Minimal Drude realization.
    (b) Corresponding normalized low-pass prototype,
    obtained from the local RLC approximation.
    (c) Generalization of the low-pass prototype through
    the insertion of a lossless two-port network
    $\widehat{\mathcal N}'$ between the series inductance
    $g_1$ and the terminating resistance $r_m$. (d)   Cauer ladder realization of a normalized low-pass
    prototype of order $N$, consisting of alternating
    series inductors and shunt capacitors, with
    normalized source resistance unity and terminating
    resistance $g_{N+1}=r_m$.
    }
\label{fig:synthesis}
\end{figure}

We first consider the minimal realization, obtained when the residual lossless two-port is the identity network [Fig. \ref{fig:synthesis}(a)].
In this case,
\begin{equation}
    Z_\chi(s)=R_m+sL_m,
    \qquad
    L_m=L_{\chi,\infty},
    \label{seq:Drude_minimal}
\end{equation}
which corresponds to a Drude susceptibility.

Consider an isolated electric-dipole mode, whose
radiation impedance
$\zeta_\Omega=R_\Omega+jX_\Omega$ is
ohmic-capacitive. For a prescribed resonance
frequency $\omega_0$, the zero-reactance condition
requires
\begin{equation}
    X_\Omega(\omega_0)+\omega_0L_m=0,
\end{equation}
and therefore
\begin{equation}
    L_m=-\frac{X_\Omega(\omega_0)}{\omega_0}.
    \label{seq:Lm_resonance}
\end{equation}
Defining $R_s:=R_\Omega(\omega_0)$, the total
driving-point impedance at resonance satisfies
\begin{equation}
    Z_\chi(j\omega_0)+\zeta_\Omega(j\omega_0)
    =
    R_s+R_m.
\end{equation}

To construct a normalized low-pass representation,
we approximate the driving-point impedance in the
vicinity of $\omega_0$.
Assuming that the frequency variation of the
radiation resistance is negligible compared with
that of the total reactance,
\begin{equation}
    |R_\Omega'(\omega_0)|
    \ll
    |X_\Omega'(\omega_0)+L_m|,
\end{equation}
the response is locally equivalent to a series
RLC circuit,
\begin{equation}
    Z_\chi(s)+\zeta_\Omega(s)
    \simeq
    R_s+R_m+sL_{\rm eq}
    +\frac{1}{sC_{\rm eq}},
    \label{seq:equivalent_RLC}
\end{equation}
where
\begin{equation}
    L_{\rm eq}
    =
    \frac{X_\Omega'(\omega_0)+L_m}{2}>0,
    \qquad
    C_{\rm eq}
    =
    \frac{1}{\omega_0^2L_{\rm eq}}.
    \label{seq:equivalent_LC}
\end{equation}
Equation~\eqref{seq:equivalent_RLC} can be rewritten as
\begin{equation}
    Z_\chi(s)+\zeta_\Omega(s)
    \simeq
    R_s
    \left[
        1+r_m+
        g_1\frac{s^2+\omega_0^2}{\beta s}
    \right],
    \label{seq:equivalent_LP}
\end{equation}
where $\beta>0$ is a frequency-scaling parameter
and
\begin{equation}
    r_m:=\frac{R_m}{R_s},
    \qquad
    g_1:=\frac{\beta L_{\rm eq}}{R_s}
\end{equation}
are dimensionless quantities.

Equation \eqref{seq:equivalent_LP} motivates the introduction
of the normalized low-pass variable
\begin{equation}
    \boxed{
    \hat{s}
    =
    \frac{s^2+\omega_0^2}{\beta s}.
    }
    \label{seq:Bp2Lp}
\end{equation}
In terms of $\hat{s}$, the normalized
driving-point impedance reduces to
\begin{equation}
    \frac{Z_\chi+\zeta_\Omega}{R_s}
    \simeq
    1+r_m+g_1\hat{s}
    \label{seq:normalized_impedance}
\end{equation}
Thus, the minimal resonant scattering problem
is represented by a normalized low-pass network
consisting of a unit source resistance,
a series inductance $g_1$, and a terminating
resistance $r_m$, as shown in Fig. \ref{fig:synthesis}(b).

On the imaginary axis, $s=j\omega$, the
frequency transformation \eqref{seq:Bp2Lp} becomes
\begin{equation}
    \hat{\omega}
    =
    \frac{\omega^2-\omega_0^2}{\beta\omega}.
    \label{seq:hat_omega}
\end{equation}
Hence, the resonance frequency $\omega_0$
is mapped to $\hat{\omega}=0$.
For a frequency band
$\mathcal B=[\omega_1,\omega_2]$ satisfying
$\omega_0^2=\omega_1\omega_2$, the band edges
are mapped to
\begin{equation}
    \hat{\omega}_1
    =
    -\frac{\Delta\omega}{\beta},
    \qquad
    \hat{\omega}_2
    =
    \frac{\Delta\omega}{\beta},
\end{equation}
where $\Delta\omega=\omega_2-\omega_1$.
Consequently, a normalized low-pass cutoff
frequency $\hat{\omega}_c$ corresponds to
\begin{equation}
    \Delta\omega=\beta\hat{\omega}_c.
    \label{seq:bandwidth_transformation}
\end{equation}
In particular, the normalization
$\hat{\omega}_c=1$ gives $\beta=\Delta\omega$.

We now return to the general Darlington
representation of the material impedance,  introducing an additional lossless two-port network
$\widehat{\mathcal N}'$ between the series
inductance $g_1$ and the terminating resistance
$r_m$ in the normalized low-pass domain, as shown in Fig. \ref{fig:synthesis}(c).

The resulting network is described by
\begin{equation}
    \hat Z_{\rm in}(\hat s)
    =
    1+g_1\hat s+
    \hat Z_{\mathcal N'}(\hat s;r_m),
    \label{seq:LP_general_network}
\end{equation}
where $\hat Z_{\mathcal N'}$ denotes the
driving-point impedance of the lossless
two-port $\widehat{\mathcal N}'$ terminated
by $r_m$.
The minimal Drude realization is recovered
when $\widehat{\mathcal N}'$ is the identity network,
so that $\hat Z_{\mathcal N'}=r_m$.
We now synthesize $\widehat{\mathcal N}'$ using
the Cauer ladder topology shown in
Fig.~\ref{fig:synthesis}(d), consisting of
alternating series inductors and shunt
capacitors. The ladder coefficients
$g_k$ determine the additional dispersive
degrees of freedom available for shaping
the absorption response.

Once the low-pass network has been synthesized,
its corresponding band-pass realization is
obtained through the frequency transformation
\begin{equation}
    \hat s
    \longmapsto
    \frac{s^2+\omega_0^2}{\beta s},
    \label{seq:Lp2Bp}
\end{equation}
following the standard methods of network
synthesis
\cite{pozar_microwave_2012,su_analog_1996}.

In particular, a series inductor of the
normalized low-pass ladder, with impedance
$\hat z_k(\hat s)=g_k\hat s$, is mapped
onto a series LC resonator:
\begin{equation}
    Z_k(s)
    =
    \frac{R_sg_k}{\beta}
    \left(
        s+\frac{\omega_0^2}{s}
    \right)
    =
    sL_k+\frac{1}{sC_k},
\end{equation}
where
\begin{equation}
    L_k=\frac{R_sg_k}{\beta},
    \qquad
    C_k=\frac{\beta}{R_sg_k\omega_0^2}.
\end{equation}
Similarly, a shunt capacitor of the
normalized low-pass ladder, with admittance
$\hat y_k(\hat s)=g_k\hat s$, is mapped
onto a parallel LC resonator:
\begin{equation}
    Y_k(s)
    =
    \frac{g_k}{R_s\beta}
    \left(
        s+\frac{\omega_0^2}{s}
    \right)
    =
    sC_k+\frac{1}{sL_k},
\end{equation}
with
\begin{equation}
    C_k=\frac{g_k}{R_s\beta},
    \qquad
    L_k=\frac{R_s\beta}{g_k\omega_0^2}.
\end{equation}
In both cases, the resulting resonators
satisfy $L_kC_k=\omega_0^{-2}$.

The frequency transformation therefore maps
the lossless Cauer ladder into a network
of series and parallel LC resonators,
all tuned to the prescribed frequency
$\omega_0$. The resulting band-pass network
provides a passive circuit realization of
the synthesized dispersive response.

In the following, we consider the Butterworth,
Chebyshev, and Bessel low-pass prototypes
realizable through the Cauer ladder topology
of Fig.~\ref{fig:synthesis} (d). Their reflection
and transmission coefficients, denoted by
$\hat\Gamma$ and $\hat t$, respectively,
satisfy $|\hat t|^2=1-|\hat\Gamma|^2$.
For notational simplicity, the hat symbol
will be omitted hereafter, with all quantities
understood in the normalized low-pass domain
unless otherwise specified.

\subsection{The Normalized Butterworth Low-Pass Response}

\begin{figure} \centering \includegraphics[width=0.85\linewidth]{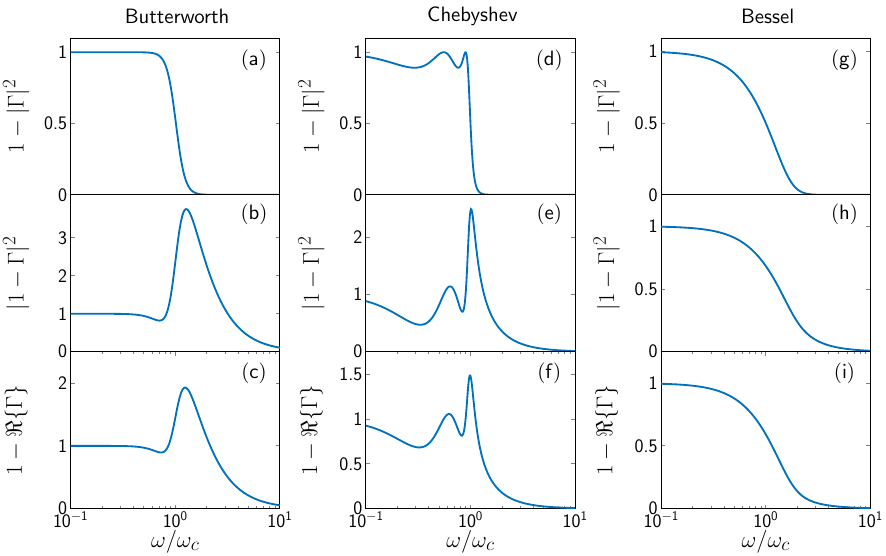} \caption{Frequency response of different linear analog filters: (a-c) Butterworth filter, (d-f) Chebyshev filter, (g-i) Bessel filter, each shown for a fifth-order filter.} \label{fig:freq_resp} \end{figure}

The Butterworth filter \cite{butterworth_theory_1930} is defined by the requirement that its magnitude response be maximally flat at zero frequency. Equivalently, for an $N$th-order realization, all derivatives of $|t(j\omega)|^2$ of orders $(1,\ldots,2N-1)$ vanish at $\omega=0$. For an $N$th-order normalized low-pass filter,
\begin{equation}
|t(j\omega)|^{2}
= \frac{1}{1 + \left({\omega}/{\omega_c}\right)^{2N}},
\label{seq:butterworth}
\end{equation}
where $\omega_c$ denotes the $-3\,\mathrm{dB}$ cutoff frequency, defined by $|t(j\omega_c)|^2 = \frac{1}{2}$. From \eqref{seq:butterworth} it follows that $\omega_c= 1$ $\forall N$.
The element values of the Cauer ladder of Fig.~\ref{fig:synthesis} (d) are then 
\begin{equation}
g_n = 2 \sin\!\left( \frac{2n-1}{2N} \pi \right),
\qquad n=1,\dots,N,
\end{equation}
with $g_{N+1}=1$ for a normalized load. The frequency response of a fifth-order Butterworth filter is shown in Fig. \ref{fig:freq_resp}(a-c).

\subsection{The Normalized Chebyshev Low-Pass Response}

The Type-I Chebyshev filter enforces an equiripple behavior in the passband, i.e., it minimizes the maximum deviation of the power transmission coefficient from unity over the passband. For an $N$th-order normalized realization,
\begin{equation}
|t(j\omega)|^2
= \frac{1}{1 + \epsilon^2 T_N^2\!\left({\omega}/{\omega_c}\right)},
\end{equation}
where { $\epsilon > 0$ is the ripple factor, $\omega_c$ is the cutoff frequency}, and $T_N$ is the Chebyshev polynomial of the first kind, defined by
\begin{equation*}
T_N(x) = \cos\left(N \cos^{-1} x\right).
\end{equation*}

The auxiliary quantities are defined as
\begin{equation*}
\beta_\epsilon = \sinh^{-1}(\epsilon^{-1}), 
\qquad
\gamma = \sinh\!\left(\frac{\beta_\epsilon}{N}\right),
\end{equation*}
\begin{equation*}
a_n = \sin\!\left(\frac{2n-1}{2N}\pi\right),
\qquad
b_n = \gamma^2 + \sin^2\!\left(\frac{n}{N}\pi\right).
\end{equation*}

The element values of the Cauer ladder of Fig. \ref{fig:synthesis}(d) are then obtained recursively as
\begin{equation}
g_1 = \frac{2 a_1}{\gamma},
\qquad
g_n = \frac{4 a_{n-1} a_n}{b_{n-1} g_{n-1}} \quad \text{with} \; n=2,3,\ldots,N
\end{equation}
with terminating element
\begin{equation}
g_{N+1} =
\begin{cases}
1, & N \text{ odd}, \\[4pt]
\coth^2(\beta_\epsilon/{2}), & N \text{ even}.
\end{cases}
\end{equation}

 The frequency response of a fifth-order Chebyshev filter is shown in Fig. \ref{fig:freq_resp}(d-f).

\subsection{The Normalized Bessel Low-Pass Response}
The Bessel filter \cite{thomson_delay_1949} is defined by the requirement of maximal flatness of the group delay $\tau_d := -\phi'(\omega)$, where $\phi = \arg  t$, at zero frequency. Equivalently, for an $N$th-order realization, all derivatives of $\tau_d (\omega)$ of orders $(1,\ldots,2N-1)$ vanish at $\omega=0$.

Unlike the Butterworth and Chebyshev responses, which optimize amplitude response, the Bessel prototype is designed to have a linear
phase response in the passband to avoid signal distortion \cite{pozar_microwave_2012}. However, this phase optimality is achieved at the expense of reduced selectivity: among classical low-pass prototypes of equal order $N$, the Bessel filter exhibits the slowest magnitude roll-off in the transition region.

For an $N$th-order Bessel filter, the transfer function is
\begin{equation}
|t(j \omega)|^2 = \left|\frac{\theta_N(0)}{\theta_N(j \omega)}\right|^2,
\end{equation}
where $\theta_N(s)$ is the reverse Bessel polynomial defined as
\begin{equation}
\theta_N(s)
=
\sum_{k=0}^{N}
\frac{(2N-k)!}{2^{\,N-k} k! (N-k)!}
\, s^{k}.
\end{equation}

The ladder coefficients $(g_n)_{n=1}^{N}$ are tabulated up to order $N=10$ in Tab. \ref{tab:Bessel_g} \cite{pozar_microwave_2012}.  The frequency response of a fifth-order Bessel filter is shown in Fig. \ref{fig:freq_resp}(g-i).

\begin{table}[h!]
\centering
\caption{Element Values for Bessel Low-Pass Filter Prototypes ($N=1,\dots,10 $) \cite{pozar_microwave_2012}}
\label{tab:Bessel_g}
\begin{tabular}{c ccccccccccc}
\hline
\hline
$N$ & $g_1$ & $g_2$ & $g_3$ & $g_4$ & $g_5$ & $g_6$ & $g_7$ & $g_8$ & $g_9$ & $g_{10}$ & $g_{11}$ \\
\hline
\hline
1  & 2.0000 & 1.0000 \\
2  & 1.5774 & 0.4226 & 1.0000 \\
3  & 1.2550 & 0.5528 & 0.1922 & 1.0000 \\
4  & 1.0598 & 0.5116 & 0.3181 & 0.1104 & 1.0000 \\
5  & 0.9303 & 0.4577 & 0.3312 & 0.2090 & 0.0718 & 1.0000 \\
6  & 0.8377 & 0.4116 & 0.3158 & 0.2364 & 0.1480 & 0.0505 & 1.0000 \\
7  & 0.7677 & 0.3744 & 0.2944 & 0.2378 & 0.1778 & 0.1104 & 0.0375 & 1.0000 \\
8  & 0.7125 & 0.3446 & 0.2735 & 0.2297 & 0.1867 & 0.1387 & 0.0855 & 0.0289 & 1.0000 \\
9  & 0.6678 & 0.3203 & 0.2547 & 0.2184 & 0.1859 & 0.1506 & 0.1111 & 0.0682 & 0.0230 & 1.0000 \\
10 & 0.6305 & 0.3002 & 0.2384 & 0.2066 & 0.1808 & 0.1539 & 0.1240 & 0.0911 & 0.0557 & 0.0187 & 1.0000 \\
\hline
\hline
\end{tabular}
\end{table}


\begin{thebibliography}{52}
\bibitem[{Chao \emph{et~al.}(2022)Chao, Strekha, Kuate~Defo, Molesky, and Rodriguez}]{chao_physical_2022}
P.~Chao, B.~Strekha, R.~Kuate~Defo, S.~Molesky, and A.~W. Rodriguez, \href{https://doi.org/10.1038/s42254-022-00468-w} {Nat. Rev. Phys. \textbf{4}, 543 (2022)}.

\bibitem[{Bode(1945)}]{bode_network_1945}
H.~W. Bode, \emph{Network analysis and feedback amplifier design}, Bell Telephone Laboratories series (D. Van Nostrand Co., New York, 1945).

\bibitem[{Fano(1947)}]{fano_theoretical_1947}
R.~M. Fano, \emph{Theoretical limitations on the broadband matching of arbitrary impedances}, \href{https://dspace.mit.edu/handle/1721.1/12909} {Ph.D. thesis}, Massachusetts Institute of Technology (1947).

\bibitem[{Chu(1948)}]{chu_physical_1948}
L.~J. Chu, \href{https://doi.org/10.1063/1.1715038} {J. Appl. Phys. \textbf{19}, 1163 (1948)}.

\bibitem[{Wheeler(1947)}]{wheeler_fundamental_1947}
H.~Wheeler, \href{https://doi.org/10.1109/JRPROC.1947.226199} {Proc. IRE \textbf{35}, 1479 (1947)}.

\bibitem[{Harrington(1960)}]{harrington_effect_1960}
R.~F. Harrington, \href{https://doi.org/10.6028/jres.064d.003} {J. Res. Natl. Bur. Stand. Sect. D \textbf{64D}, 1 (1960)}.

\bibitem[{Gustafsson \emph{et~al.}(2012)Gustafsson, Cismasu, and Jonsson}]{gustafsson_physical_2012}
M.~Gustafsson, M.~Cismasu, and B.~L.~G. Jonsson, \href{https://doi.org/10.1109/TAP.2012.2194658} {IEEE Trans. Antennas Propag. \textbf{60}, 2672 (2012)}.

\bibitem[{Miller \emph{et~al.}(2014)Miller, Hsu, Reid, Qiu, DeLacy, Joannopoulos, Solja\v{c}i\'{c}, and Johnson}]{miller_fundamental_2014}
O.~D. Miller, C.~W. Hsu, M.~T.~H. Reid, W.~Qiu, B.~G. DeLacy, J.~D. Joannopoulos, M.~Solja\v{c}i\'{c}, and S.~G. Johnson, \href{https://doi.org/10.1103/PhysRevLett.112.123903} {Phys. Rev. Lett. \textbf{112}, 123903 (2014)}.

\bibitem[{Miller \emph{et~al.}(2016)Miller, Polimeridis, Reid, Hsu, DeLacy, Joannopoulos, Solja\v{c}i\'{c}, and Johnson}]{miller_fundamental_2016}
O.~D. Miller, A.~G. Polimeridis, M.~T.~H. Reid, C.~W. Hsu, B.~G. DeLacy, J.~D. Joannopoulos, M.~Solja\v{c}i\'{c}, and S.~G. Johnson, \href{https://doi.org/10.1364/OE.24.003329} {Opt. Express \textbf{24}, 3329 (2016)}.

\bibitem[{Kuang \emph{et~al.}(2020)Kuang, Zhang, and Miller}]{kuang_maximal_2020}
Z.~Kuang, L.~Zhang, and O.~D. Miller, \href{https://doi.org/10.1364/OPTICA.398715} {Optica \textbf{7}, 1746 (2020)}.

\bibitem[{Sohl \emph{et~al.}(2007{a})Sohl, Gustafsson, and Kristensson}]{sohl_physical_2007}
C.~Sohl, M.~Gustafsson, and G.~Kristensson, \href{https://doi.org/10.1088/0022-3727/40/22/042} {J. Phys. D: Appl. Phys. \textbf{40}, 7146 (2007{a})}.

\bibitem[{Sohl \emph{et~al.}(2007{b})Sohl, Gustafsson, and Kristensson}]{sohl_physical_ostacles_2007}
C.~Sohl, M.~Gustafsson, and G.~Kristensson, \href{https://doi.org/10.1088/1751-8113/40/36/015} {J. Phys. A: Math. Theor. \textbf{40}, 11165 (2007{b})}.

\bibitem[{Molesky \emph{et~al.}(2020)Molesky, Chao, Jin, and Rodriguez}]{molesky_global_2020}
S.~Molesky, P.~Chao, W.~Jin, and A.~W. Rodriguez, \href{https://doi.org/10.1103/PhysRevResearch.2.033172} {Phys. Rev. Research \textbf{2}, 033172 (2020)}.

\bibitem[{Gustafsson \emph{et~al.}(2020)Gustafsson, Schab, Jelinek, and Capek}]{gustafsson_upper_2020}
M.~Gustafsson, K.~Schab, L.~Jelinek, and M.~Capek, \href{https://doi.org/10.1088/1367-2630/ab83d3} {New J. Phys. \textbf{22}, 073013 (2020)}.

\bibitem[{Shim \emph{et~al.}(2019)Shim, Fan, Johnson, and Miller}]{shim_fundamental_2019}
H.~Shim, L.~Fan, S.~G. Johnson, and O.~D. Miller, \href{https://doi.org/10.1103/PhysRevX.9.011043} {Phys. Rev. X \textbf{9}, 011043 (2019)}.

\bibitem[{Rozanov(2000)}]{rozanov_ultimate_2000}
K.~Rozanov, \href{https://doi.org/10.1109/8.884491} {IEEE Trans. Antennas Propag. \textbf{48}, 1230 (2000)}.

\bibitem[{Monticone and Al\`{u}(2016)}]{monticone_invisibility_2016}
F.~Monticone and A.~Al\`{u}, \href{https://doi.org/10.1364/OPTICA.3.000718} {Optica \textbf{3}, 718 (2016)}.

\bibitem[{Sup()}]{SuppMat}
See the Supplemental Material for the derivation of the Bode--Fano limit, which includes Refs.~\cite{shlivinski_beyond_2018,forestiere_volume_2018,forestiere_material-independent_2016,mayergoyz_electrostatic_2005,forestiere_resonance_2020}.

\bibitem[{Forestiere \emph{et~al.}(2024)Forestiere, Miano, and Al\`{u}}]{forestiere_first-principles_2024}
C.~Forestiere, G.~Miano, and A.~Al\`{u}, \href{https://doi.org/10.1103/PhysRevApplied.22.034014} {Phys. Rev. Applied \textbf{22}, 034014 (2024)}.

\bibitem[{Hanson and Yakovlev(2002)}]{hanson_operator_2002}
G.~W. Hanson and A.~B. Yakovlev, \href{https://doi.org/10.1007/978-1-4757-3679-3} {\emph{Operator Theory for Electromagnetics}} (Springer, New York, NY, 2002).

\bibitem[{Mishchenko(2014)}]{mishchenko_electromagnetic_2014}
M.~I. Mishchenko, \href{https://doi.org/10.1017/CBO9781139019064} {\emph{Electromagnetic Scattering by Particles and Particle Groups: An Introduction}} (Cambridge University Press, Cambridge, 2014).

\bibitem[{Corsaro \emph{et~al.}(2026)Corsaro, Balato, Miano, Petrarca, Al\`u, and Forestiere}]{corsaro_mie_2026}
E.~Corsaro, M.~Balato, G.~Miano, C.~Petrarca, A.~Al\`u, and C.~Forestiere, \href{https://doi.org/10.1103/n9kj-6j67} {Phys. Rev. Lett. \textbf{136}, 223802 (2026)}.

\bibitem[{Brune(1931)}]{brune_synthesis_1931}
O.~Brune, \href{https://doi.org/10.1002/sapm1931101191} {J. Math. Phys. \textbf{10}, 191 (1931)}.

\bibitem[{Cauer(1932)}]{cauer_poisson_1932}
W.~Cauer, \href{https://doi.org/10.1090/S0002-9904-1932-05510-0} {Bulletin of the American Mathematical Society \textbf{38}, 713 (1932)}.

\bibitem[{Gustafsson(2010)}]{gustafsson_sum_2010}
M.~Gustafsson, \href{https://doi.org/10.1049/iet-map.2009.0083} {IET Microw. Antennas Propag. \textbf{4}, 501 (2010)}.

\bibitem[{Altarelli \emph{et~al.}(1972)Altarelli, Dexter, Nussenzveig, and Smith}]{altarelli_superconvergence_1972}
M.~Altarelli, D.~L. Dexter, H.~M. Nussenzveig, and D.~Y. Smith, \href{https://doi.org/10.1103/PhysRevB.6.4502} {Physical Review B \textbf{6}, 4502 (1972)}.

\bibitem[{Purcell(1969)}]{purcell_absorption_1969}
E.~M. Purcell, \href{https://doi.org/10.1086/150207} {Astrophys. J. \textbf{158}, 433 (1969)}.

\bibitem[{Bohren and Huffman(1998)}]{bohren_absorption_1998}
C.~F. Bohren and D.~R. Huffman, \href{https://doi.org/10.1002/9783527618156} {\emph{Absorption and Scattering of Light by Small Particles}} (Wiley, New York, 1998).

\bibitem[{Mishchenko(2008)}]{mishchenko_broadband_2008}
M.~I. Mishchenko, \href{https://doi.org/10.1364/JOSAA.25.002893} {J. Opt. Soc. Am. A \textbf{25}, 2893 (2008)}.

\bibitem[{Youla(1964)}]{youla_new_1964}
D.~C. Youla, \href{https://doi.org/10.1109/TCT.1964.1082267} {IEEE Transactions on Circuit Theory \textbf{11}, 30 (1964)}.

\bibitem[{Pozar(2012)}]{pozar_microwave_2012}
D.~M. Pozar, \emph{Microwave Engineering}, 4th ed. (Wiley, Hoboken, NJ, 2012).

\bibitem[{Pascale \emph{et~al.}(2023)Pascale, Mann, Tzarouchis, Miano, Alu, and Forestiere}]{pascale_lower_2023}
M.~Pascale, S.~A. Mann, D.~C. Tzarouchis, G.~Miano, A.~Alu, and C.~Forestiere, \href{https://doi.org/10.1109/TAP.2023.3248442} {IEEE Transactions on Antennas and Propagation \textbf{71}, 4350 (2023)}.

\bibitem[{Darlington(1939)}]{darlington_synthesis_1939}
S.~Darlington, \href{https://doi.org/10.1002/sapm1939181257} {J. Math. Phys. \textbf{18}, 257 (1939)}.

\bibitem[{Su(1996)}]{su_analog_1996}
K.~L. Su, \href{https://doi.org/10.1007/978-1-4615-2051-1} {\emph{Analog Filters}} (Springer US, Boston, MA, 1996).

\bibitem[{Van~Valkenburg(1974)}]{van_valkenburg_network_1974}
M.~E. Van~Valkenburg, \emph{Network analysis}, 3rd ed. (Prentice-Hall, Englewood Cliffs, N.J, 1974).

\bibitem[{Butterworth(1930)}]{butterworth_theory_1930}
S.~Butterworth, Experimental Wireless \& the Wireless Engineer \textbf{7}, 536 (1930).

\bibitem[{Thomson(1949)}]{thomson_delay_1949}
W.~E. Thomson, \href{https://doi.org/10.1049/pi-3.1949.0101} {Proc. IEE Part III \textbf{96}, 487 (1949)}.

\bibitem[{Matthaei(1956)}]{matthaei_synthesis_1956}
G.~Matthaei, \href{https://doi.org/10.1109/TCT.1956.1086315} {IRE Transactions on Circuit Theory \textbf{3}, 163 (1956)}.

\bibitem[{Kelly \emph{et~al.}(2003)Kelly, Coronado, Zhao, and Schatz}]{kelly_optical_2003}
K.~L. Kelly, E.~Coronado, L.~L. Zhao, and G.~C. Schatz, \href{https://doi.org/10.1021/jp026731y} {Journal of Physical Chemistry B \textbf{107}, 668 (2003)}.

\bibitem[{Mie(1908)}]{mie_beitrage_1908}
G.~Mie, \href{https://doi.org/10.1002/andp.19083300302} {Ann. Phys. (Berlin) \textbf{330}, 377 (1908)}.

\bibitem[{Han and Park(2023)}]{Han23}
W.~Han and Q.-H. Park, \href{https://doi.org/10.1515/nanoph-2022-0777} {Nanophotonics \textbf{12}, 2443 (2023)}.

\bibitem[{Li \emph{et~al.}(2021)Li, Lipton, and Maier}]{li_lorentz_2021}
W.~Li, R.~Lipton, and M.~Maier, \href{https://doi.org/10.1098/rspa.2021.0609} {Proceedings of the Royal Society A: Mathematical, Physical and Engineering Sciences \textbf{477}, 20210609 (2021)}.

\bibitem[{Lee \emph{et~al.}(2014)Lee, Tymchenko, Argyropoulos, Chen, Lu, Demmerle, Boehm, Amann, Al\`{u}, and Belkin}]{lee_giant_2014}
J.~Lee, M.~Tymchenko, C.~Argyropoulos, P.-Y. Chen, F.~Lu, F.~Demmerle, G.~Boehm, M.-C. Amann, A.~Al\`{u}, and M.~A. Belkin, \href{https://doi.org/10.1038/nature13455} {Nature \textbf{511}, 65 (2014)}.

\bibitem[{Mann \emph{et~al.}(2021)Mann, Nookala, Johnson, Cotrufo, Mekawy, Klem, Brener, Raschke, Al\`{u}, and Belkin}]{mann_ultrafast_2021}
S.~A. Mann, N.~Nookala, S.~C. Johnson, M.~Cotrufo, A.~Mekawy, J.~F. Klem, I.~Brener, M.~B. Raschke, A.~Al\`{u}, and M.~A. Belkin, \href{https://doi.org/10.1364/OPTICA.415581} {Optica \textbf{8}, 606 (2021)}.

\bibitem[{Al{\`u}(2011)}]{Alu11}
A.~Al{\`u}, \href{https://doi.org/10.1103/PhysRevB.84.075153} {Phys. Rev. B \textbf{84}, 075153 (2011)}.

\bibitem[{Li \emph{et~al.}(2019)Li, Mekawy, and Al\`{u}}]{li_beyond_2019}
H.~Li, A.~Mekawy, and A.~Al\`{u}, \href{https://doi.org/10.1103/PhysRevLett.123.164102} {Phys. Rev. Lett. \textbf{123}, 164102 (2019)}.

\bibitem[{Hayran and Monticone(2024)}]{hayran_beyond_2024}
Z.~Hayran and F.~Monticone, \href{https://doi.org/10.1103/PhysRevApplied.21.044007} {Phys. Rev. Applied \textbf{21}, 044007 (2024)}.

\bibitem[{Shlivinski and Hadad(2018)}]{shlivinski_beyond_2018}
A.~Shlivinski and Y.~Hadad, \href{https://doi.org/10.1103/PhysRevLett.121.204301} {Phys. Rev. Lett. \textbf{121}, 204301 (2018)}.

\bibitem[{Forestiere \emph{et~al.}(2018)Forestiere, Miano, Rubinacci, Tamburrino, Tricarico, and Ventre}]{forestiere_volume_2018}
C.~Forestiere, G.~Miano, G.~Rubinacci, A.~Tamburrino, R.~Tricarico, and S.~Ventre, IEEE Transactions on Antennas and Propagation \textbf{66}, 2505 (2018).

\bibitem[{Forestiere and Miano(2016)}]{forestiere_material-independent_2016}
C.~Forestiere and G.~Miano, Physical Review B \textbf{94}, 201406 (2016).

\bibitem[{Mayergoyz \emph{et~al.}(2005)Mayergoyz, Fredkin, and Zhang}]{mayergoyz_electrostatic_2005}
I.~D. Mayergoyz, D.~R. Fredkin, and Z.~Zhang, \href{https://doi.org/10.1103/PhysRevB.72.155412} {Phys. Rev. B \textbf{72}, 155412 (2005)}.

\bibitem[{Forestiere \emph{et~al.}(2020)Forestiere, Miano, and Rubinacci}]{forestiere_resonance_2020}
C.~Forestiere, G.~Miano, and G.~Rubinacci, \href{https://doi.org/10.1103/PhysRevResearch.2.043176} {Physical Review Research \textbf{2}, 043176 (2020)}.

\end{thebibliography}
\end{document}